\documentclass[aip,jcp]{revtex4-2}
\usepackage{graphicx,amssymb,soul,color,amsmath,bm}
\usepackage{epstopdf,hyperref}
\usepackage{braket,dsfont,nicefrac}
\usepackage[mathcal]{euscript}
\usepackage[normalem]{ulem}
\usepackage[utf8]{inputenc}

\usepackage{algorithm}
\usepackage{ytableau}
\usepackage{algpseudocode}
\usepackage{array}
\usepackage{tabularx}

\usepackage{multirow}
\newcolumntype{C}{>{\centering\arraybackslash}X}

\makeatletter
\def\@email#1#2{%
 \endgroup
 \patchcmd{\titleblock@produce}
  {\frontmatter@RRAPformat}
  {\frontmatter@RRAPformat{\produce@RRAP{*#1\href{mailto:#2}{#2}}}\frontmatter@RRAPformat}
  {}{}
}%

\hypersetup{colorlinks=true,citecolor=blue,linkcolor=magenta}

\sethlcolor{yellow}
\allowdisplaybreaks

\begin{document}

%\title{t-J chain with long range spin couplings and Coulomb interactions}
\title{Variational Block Correlated Coupled Cluster Monte Carlo}

\author{Hunter Brodie}
\affiliation{Rudolf Peierls Centre for Theoretical Physics, University of Oxford, Oxford OX1 3NP, United Kingdom}
\affiliation{Department of Physics, Northeastern University, Boston, Massachusetts 02115, USA}

\author{Alexander Shilcusky}
\affiliation{Department of Physics, Northeastern University, Boston, Massachusetts 02115, USA}
%\author{Anders Sandvik}
%\affiliation{Department of Physics, Boston University, Boston, Massachusetts 02215, USA}
\author{Adrian Feiguin}
\email{a.feiguin@northeastern.edu}
\affiliation{Department of Physics, Northeastern University, Boston, Massachusetts 02115, USA}

\begin{abstract}
%The recent success of quantum neural states has renewed interest in physics-informed variational wave functions for quantum many-body systems. 
Motivated by the recent success of quantum neural states, we present a variational implementation of the coupled cluster method suitable for quantum Monte Carlo simulations. We introduce a Young tableaux representation of the wave-function coefficients and their derivatives that substantially reduces the computational complexity of the variational optimization. While the algorithm is generic, we illustrate its implementation with an application to the Heisenberg model in one and two dimensions, using the  block correlated coupled cluster ansatz (BCCC) with doubles and triples. Convergence is greatly improved by means of importance sampling based on a fictitious temperature. The resulting approach achieves competitive accuracy while providing a scalable framework for variational coupled-cluster calculations.  
\end{abstract}
\maketitle

\section{Introduction}

Understanding correlated quantum systems requires going beyond the single-particle picture and confronting the many-body problem directly. The resulting configuration space grows exponentially with system size, making an exact representation impossible even for the largest supercomputers. This is a pervasive problem in nuclear physics, quantum chemistry, and condensed matter, and there is an urgent need for unbiased and scalable numerical methods that can produce reliable results beyond one-dimension.

In recent years we have witnessed impressive and very promising progress based on the use of neural network (NN) wave functions as variational estimators \cite{Carleo2017,Saito2017,Cai2018,Glasser2018,Medvidovic2024,Lange2024}. The power of ``neural quantum states'' (NQS) lies in the complex non-linear structure that provides them with remarkable expressivity to represent arbitrary complex many-body states, even though they are completely agnostic to the physics.  Variational Monte Carlo with neural quantum states has demonstrated great promise, becoming one of the most active areas of research in computational condensed matter and chemistry. 

These developments have stimulated the search for improved optimization algorithms and more expressive variational wave functions. Despite their remarkable flexibility, most neural quantum states lack an explicit physics-informed architecture. Consequently, identifying an effective network architecture for a given problem remains largely an empirical task, relying on trial and error and computationally demanding optimization over thousands or even millions of variational parameters.\cite{Medvidovic2024}.

One such approach is to use the so-called vision transformers to represent the ground-state of quantum spin systems \cite{viteritti2023vit,Sobral2025}. The idea consists of subdividing the lattice into blocks of finite size $N_0$, such that each spin configuration inside the blocks is treated as a ``word'' in a ``sentence'', whose correlations are learned by a prescribed attention mechanism. \cite{rende2025queries}
%For spin-1/2 problems, it has been shown a relative positional encoding works better than the traditional softmax ``keys and queries" mechanism .

This motivates us to explore other variational ansatze that also rely in partitioning the system into blocks, with the aim of gaining insight into the structure of the correlations to design more advanced physics-informed neural quantum states. Approaches that have experienced certain success and hold a great deal of promise, include the use of ``entangled plaquette states'' \cite{Mezzacapo2009}, correlator product state wave-functions\cite{Changlani2009,Neuscamman2011} and beyond-mean-field approaches \cite{Isaev2009,Kastaros2022}. In this work, we focus on the ``block correlated coupled cluster'' (BCCC) method\cite{Li2004,Wang2020,ren2024block}, a technique that combines ideas from quantum information and electronic structure calculations. 

Although the ideas described in this work are broadly applicable, we use the antiferromagnetic Heisenberg Hamiltonian as a testbed since highly accurate results are available both in one dimension and on the square lattice. It is defined as:
\begin{equation}
H_{XXZ}=\sum_{\langle ij \rangle} \Delta S_i^zS_j^z+\frac{1}{2}\left(S_i^+S_j^-+S_i^-S_j^+\right),
\end{equation}
where we have introduced an Ising anisotropy $\Delta$. The spin operators $S^z_i$ act on the localized spin positioned at site $i$ (we consider $S=1/2$ in this work), $S^+$ and $S^-$ are the raising and lowering operators, and the summation runs over nearest neighbors. The physics of this model is well established. In one dimension, it can be solved exactly using the Bethe ansatz \cite{Bethe1031,integrability}, revealing quasi-long-range antiferromagnetic order characterized by algebraically decaying correlations for $\Delta=1$, and a gapped phase with a finite correlation length for $\Delta>1$. In addition, density matrix renormalization group (DMRG) calculations \cite{White1992,White1993,Schollwock2005,Schollwock2011,Feiguin2013a} provide highly accurate benchmark results for finite systems. On the square lattice, sign-problem-free quantum Monte Carlo simulations for \cite{Sandvik1997} enable accurate numerical studies and demonstrate that the model exhibits spontaneous symmetry breaking with true long-range antiferromagnetic order for $\Delta=1$.

This paper is organized as follows: In Sec.\ref{sec:CC} we describe coupled-cluster theory, focusing on the BCCC approach. In Sec.\ref{sec:VMC} we introduce the variational implementation, including the evaluation of the wave function amplitudes, the variational optimization, and importance sampling. In Sec.\ref{sec:results} we illustrate the application to the Heisenberg model. We close with a discussion and perspectives for further developments.

\section{Coupled Cluster theory}\label{sec:CC}
Originally developed in the context of nuclear many-body theory \cite{Kummel2003,Hagen2010,Hagen2012,Horoi2007,Sun2023}, coupled-cluster theory has evolved into one of the most successful first-principles approaches in quantum chemistry, where it is widely regarded as a benchmark method for treating electron correlation \cite{Barlett2007,Lyakh2012}. In addition to these fields, it also found applications in condensed matter\cite{Zeng1998,Bishop2000,Roger1990,Zinke2008,Darradi2008,Richter2010,Richter2015,Bishop2019,Papas2022}. The main concept relies on describing a quantum many body state as a linear combination of ``fluctuations'' on top of a reference state, such as particle-hole excitations above a non-interacting Fermi sea. In block-correlated coupled cluster theory (BCCC) \cite{Li2004}, a lattice is partitioned into a periodic array of clusters (or blocks/plaquettes), in a similar spirit as renormalization group approaches. 
Each cluster/block contains a number $N_0$ of sites, or spins in the case of a spin Hamiltonian. Instead of working in Fock space ($S^z$ basis), we diagonalize the cluster Hamiltonian and use its eigenvectors as the block basis (other approaches are possible, see Ref.\onlinecite{Li2004}). It is convenient to classify them according to their energy and symmetries (quantum numbers) of each state ($E$,$S$,$S^z$, reflection symmetry, rotations...), that identify them uniquely (in principle). States that are related by time-reversal, for example, will have the same energy, but opposite $S^z$. 

The zero'th order approximation to the ground state will be a trivial product state $|\Phi_0\rangle=|00\cdots 0\rangle$, where $|0\rangle$ is the cluster ground state, or vacuum. 
%In general, we can impose the property that the total $S^z_{tot}=0$. While sampling, states with lower energy have larger probability or appearing, but states with higher energy  have larger probability to appear farther apart or alternating with one another. 
%For example, in the case of spins, we can choose the reference state $|\Phi_0\rangle=|00\cdots 0\rangle$, where $|0\rangle \equiv |s\rangle$, the ground state of a block. 
One can then consider a basis of single excitations
$|\Phi^{i_I}\rangle=|00\cdots i_I \cdots 00\rangle$, double excitations $|\Phi^{i_Ij_J}\rangle = |00\cdots i_I \cdots j_J \cdots 00\rangle$, etc, where the index refers to a basis state $|i\rangle$ at block position $I$. An generic many-body state can be written as
\[
|\psi\rangle = \sum_{i(I)}\sum_{j(J)}\cdots \sum_{m(M)}\psi_{i_Ij_J\cdots m_M} |i_Ij_J\cdots m_M\rangle.
\]
We can separate the contributions according to the number of excitations:
\begin{eqnarray}
|\Phi_0\rangle &=& |00\cdots 0\rangle \nonumber \\ %\equiv |ss \cdots s\rangle, \nonumber \\
|\Phi^{i_I}\rangle &=& |0\cdots 0 i_I 0 \cdots 0\rangle = (|i_I\rangle\langle 0_I|)|\Phi_0\rangle \nonumber \\
|\Phi^{i_Ij_J}\rangle 
&=& |0\cdots 0 i_I 0 \cdots 0 j_J 0 \cdots 0\rangle \nonumber \\
&=& (|j_J\rangle\langle 0_J|)(|i_I\rangle\langle 0_I|)|\Phi_0\rangle \nonumber \\
&=& |j_J\rangle |i_I\rangle \langle 0_J|\langle 0_I|\Phi_0\rangle, \nonumber 
\label{phis}
\end{eqnarray}
and so on. Accordingly, the many body state can be reformulated as
\begin{eqnarray}
|\Psi\rangle &=& \Psi_0|\Phi_0\rangle + \sum_I \sum_{i \in I} C^{i_I}|\Phi^{i_I}\rangle + \nonumber \\
&+&\sum_I \sum_{J > I} \sum_{i \in I} \sum_{j \in J} C^{i_Ij_J}|\Phi^{i_Ij_J}\rangle \nonumber \\
&+& \sum_I \sum_{J > I} \sum_{K > J}\sum_{i \in I} \sum_{j \in J} \sum_{k \in K}C^{i_Ij_Jk_K}|\Phi^{i_Ij_Jk_K}\rangle + \cdots, \nonumber
\end{eqnarray}
where we have assumed that the sums do not include the cluster ground state $|0_I\rangle$.
 The ground state ansatz is expressed as
 \begin{equation}
 |\Psi\rangle = e^{T}|\Phi_0\rangle = (1+T+\frac{T^2}{2!}+\frac{T^3}{3!}+\cdots)|\Phi_0\rangle=\sum_{n}\frac{T^n}{n!}|\Phi_0\rangle,
 \label{CC}
 \end{equation}
 with $T=T_1+T_2+T_3+\cdots$. By introducing the excitation operators (or ``excitors'') $A_i^I=|i_I\rangle\langle0_I|$, the $n$-block correlation operators are defined as:
% \begin{eqnarray}
%T_1& = &\sum_I \sum_{i \in I} t^{i_I} |i_I\rangle\langle 0_I|\Phi_0\rangle \nonumber \\
%T_2 &=& \frac{1}{2!}\sum_I \sum_{J > I} \sum_{i \in I} \sum_{j \in J} t^{i_Ij_J} |j_J\rangle |i_I\rangle \langle 0_J|\langle 0_I|\Phi_0\rangle, \nonumber \\
%T_3 &=& \frac{1}{3!}\sum_I \sum_{J > I} \sum_{K > J}\sum_{i \in I} \sum_{j \in J} \sum_{k \in K} t^{i_Ij_Jk_K} |k_K\rangle |j_J\rangle |i_I\rangle \langle 0_K|\langle 0_J|\langle 0_I|\Phi_0\rangle, \nonumber
% \end{eqnarray}
 \begin{eqnarray}
T_1& = &\sum_I \sum_{i \in I} t_i^I A_i^I|\Phi_0\rangle \nonumber \\
T_2 &=& \sum_I \sum_{J > I} \sum_{i \in I} \sum_{j \in J} t_{ij}^{IJ} |j_J\rangle A_i^IA_j^J|\Phi_0\rangle, \nonumber \\
T_3 &=& \sum_I \sum_{J > I} \sum_{K > J}\sum_{i \in I} \sum_{j \in J} \sum_{k \in K} t_{ijk}^{IJK} A_i^IA_j^JA_k^K|\Phi_0\rangle, \nonumber
 \end{eqnarray}
 and similar higher order contributions. A truncation to order $n$ is referred-to as BCCC$n$ approximation. In these expressions, the excitation amplitudes $t_i^{I},t_{ij}^{IJ},t_{ijk}^{IJK}$, etc. are known as as ``singles'',``doubles'',``triples'' and so on (BCCC2 would contain only singles and doubles, and BCCC3 would also include triples, for instance). They preserve the symmetries of the lattice ({\it e.g.} in a lattice model with translational invariance, $t_{ij}^{IJ}$ will only depend on the relative positions of 
 blocks $I$ and $J$) but can be antisymmetric with respect of an exchange of $i$ and $j$.  
 Notice also that the excitation operators are not Hermitian, commute with one another if acting on separate blocks, and that $(A_i^I)^2=0$. We also point out that we assume that all excitations should preserve $U(1)$ symmetry, such as total particle number or spin $S^z$. 

In order to carry out a calculation, we need to know the projections $\langle x|\Psi\rangle$ on an arbitrary basis state $|\Phi\rangle=|i_ij_J\cdots m_M\rangle$. We can proceed by evaluating each $n$-tuply excited state as
\begin{eqnarray}
\langle \Phi^{i_I}|\Psi\rangle &=& \langle \Phi^{iI}|T_1|\Phi_0\rangle = t_i^I \nonumber \\
\langle \Phi^{i_Ij_J}|\Psi\rangle &=& \langle \Phi^{i_Ij_J}|\frac{1}{2!}T_1^2+T_2|\Phi_0\rangle = t_i^{I}t_j^{J}+t_{ij}^{IJ}  \\ 
\langle\Phi^{i_Ij_Jk_K}|\Psi\rangle &=&\langle\Phi^{i_Ij_Jk_K}|\frac{1}{3!}T_1^3+\frac{1}{2!}(T_1T_2+T_2T_1)+T_3|\Phi_0\rangle \nonumber \\  &=&t_i^{I}t_j^{J}t_k^{K}+t_{ij}^{IJ}t_k^{K}+t_{ik}^{IK}t_j^{J}+t_{jk}^{JK}t_i^{I}+t_{ijk}^{IJK}, \nonumber
\label{t_eqs}
\end{eqnarray}
and so on.

%As we can infer from these expressions, at a given level of truncation $l$ of the exponential, the same operator $T_q$ can appear with a given power $(T_q)^m$, with $m<l$, which is referred to as the ``cluster size''-- hence the moniker ``coupled cluster''. 
As the level of truncation $l$ increases, the number of possible combinations appearing in these expressions grows exponentially, which requires efficient calculation schemes\cite{Zeng1998,Thom2010,Scott2017,Spencer2016}. Notice that the cluster amplitudes $t$ appear in products and powers with other amplitudes, leading to non-linear equations for the energy estimation. 

The conventional method to optimize the CC amplitudes $t$ consists of utilizing a projective technique that starts from the assumption that the CC wave-function can provide a faithful representation of the ground state, such that:
\begin{eqnarray}
H|\Psi\rangle = E|\Psi\rangle \Rightarrow \langle \Phi_0 |H|\Psi\rangle = E.
\label{projection}
\end{eqnarray}
where we have implicitly taken $\langle \Phi_0|\Psi\rangle = 1$. We can generalize this procedure by considering the projections on other basis states:
\begin{eqnarray}
\langle \Phi^{i_I} |He^T|\Phi_0\rangle &=&E\langle\Phi^{i_I}|\Psi\rangle = Et_i^I\nonumber \\
\langle \Phi^{i_Ij_J} |He^T|\Phi_0\rangle &=&E\langle\Phi^{i_Ij_J}|\Psi\rangle = E(t_i^It_j^I+t_{ij}^{IJ}) 
\end{eqnarray}
where we have inserted the expressions (\ref{t_eqs}) to obtain a system of equations for $E$ and the amplitudes $t$ that can be solved iteratively.

%We propose to use a projection technique reminiscent of natural gradient descent methods used for optimizing NNQS\cite{SorellaBook}. The idea is to systematically improve the wave function by evolving the state in imaginary time. %At each step, we obtain a new state as $|\Psi'\rangle = (1-\tau H)|\Psi\rangle$, where $\tau$ is a small parameter that can be associated to a ``learning rate''. %To lowest order, this method yields a set of equations\cite{Thom2010}:
%\begin{eqnarray}
%{t_i^I}' &=& t_i^I-E_i^I t_i^I \nonumber \\
%{t_{ij}^{IJ}}'&=&t_{ij}^{IJ}- \langle\Phi^{i_Ij_J}|H|\Psi\rangle, \\
%{t_i^I}'{t_j^J}'+{t_{ij}^{IJ}}' &=& t_i^I t_j^J+t_{ij}^{IJ} \nonumber \\ &-& \sum_{KL}\sum_{kl}\langle\Phi^{k_Kl_L}|H|\Phi^{i_Ij_J}\rangle (t_k^K t_l^L+t_{kl}^{KL}),\cdots \nonumber
\label{minimization}
%\end{eqnarray}

\section{Variational CC Monte Carlo}\label{sec:VMC}

The projective technique described above is a simple and economical approximation to a many body wave function in which the reference state is a single Slater determinant. This is because, in electronic structure calculations, Slater's rules state that the matrix elements between two determinants that differ by more than two orbitals are zero. Hence, a truncation to second order is the ``natural'' choice in such a case\cite{Crawford2000}. Moreover, double excitations capture most of the correlations, even for spin Hamiltonians. Higher order corrections can be introduced perturbatively\cite{RAGHAVACHARI1989,Papas2022}, since the computational cost of going beyond doubles becomes increasingly prohibitive\cite{Noga1987}. One important aspect that one should notice is that, since the results are not variational, they are not necessarily providing an upper bound to the energy, implying that this estimate cannot be used to control the quality of the approximation\cite{Neuscamman2011}. At this point, it is worth considering a legitimate variational approach instead, where the energy is obtained as\cite{Mezzacapo2009}: 
%To this aim, we need to introduce the wave function normalization explicitly in the definition, Eq.(\ref{CC}), and sample the expectation values and overlaps in Eq.(\ref{minimization}) according to  
\begin{equation}
%\langle \Phi|H|\Psi\rangle= |\langle
%\Psi|\Psi\rangle|^2 \frac{\sum_x E_x |\langle x|\Psi\rangle|^2}{\sum_x |\langle x|\Psi\rangle|^2}, \,\,\mathrm{with} \,\, E_x=\frac{\langle\Phi|H|x\rangle}{\langle \Psi|x\rangle}.
%\end{equation}
\frac{\langle \Psi|H|\Psi\rangle}{\langle \Psi|\Psi\rangle}= \frac{\sum_x E_x |\langle x|\Psi\rangle|^2}{\sum_x |\langle x|\Psi\rangle|^2}=\sum_x P_x E_x 
\label{Ev}
\end{equation} 
with
\[
P_x=\frac{p_z}{\sum_z p_x}; p_x= |\langle x|\Psi\rangle|^2; E_x=\frac{\langle\Psi|H|x\rangle}{\langle \Psi|x\rangle}.
\]
In these expressions, the states $|x\rangle$ represent the basis states $|\Phi_0\rangle,|\Phi^{I_i}\rangle,|\Phi^{I_iJ_j}\rangle\cdots$. 
The general argument against such approach is that one cannot truncate to second order, even in the case of Slater determinants, and hence one needs to account for higher order terms as well\cite{Crawford2000,ARPONEN1983311,BARTLETT198829}.  
One possible solution is to utilize a Monte Carlo approach to sample the contributions to (\ref{Ev}).
Previous stochastic methods\cite{Thom2010,Scott2017,Spencer2016} resemble diffusion Monte Carlo and rely on an evolving population of random walkers that dynamically ``spawn'' and ``die''. In this work we propose a variational approach that relies on the smart use of data structures and an original diagrammatic representation of the wave function coefficients in terms of the coupled cluster amplitudes that allows one to encode all the possible products of the expansion, and at the same time enabling very fast updates.

The basic premise consists of sampling according to a probability distribution $P_x$.
The random walk is carried out following a ``Hamiltonian update'' scheme 

%: given a state $|x\rangle$, a new candidate move is chosen among the states $|y\rangle$ generated by the application of $H$ to $|x\rangle$,
%\begin{equation}
%H|x\rangle = \sum_y \langle y|H|x\rangle |y\rangle.
%\label{Hx}
%\end{equation}
The Hamiltonian is built in the block eigenbasis by accounting for all interactions between blocks:
\begin{equation}
H=\sum_i H_{0_I}+\sum_{I\neq J} H_{IJ}.
\label{Hx}
\end{equation}
%Since the inter-block terms $H_{IJ}$ are not necessarily sparse, this introduces some overhead. 
Given two neighboring plaquettes in states $|i_Ij_J\rangle$, a new candidate Monte Carlo move $|i_I'j_J'\rangle$ is chosen with uniform probability among all those states connected to $|i_Ij_J\rangle$ through the action of $H_{IJ}$ (Notice that, in addition, the Hamiltonian may also act on the reference state $|0\rangle$ through the excitors). 
Since $H_{IJ}$ consists of two-body terms connecting two neighboring blocks, one only needs to store relatively small $D\times D$ matrices, where $D$ is the local dimension of the block basis ($D=2^{N_0}$ for the spin $S=1/2$ Heisenberg model). The transition probability between two configurations $|x\rangle$ and a new one $|y\rangle$ is given by $w=p_y/p_x$, which is positive definite and does not require the evaluation of the norm of $|\Psi\rangle$.

\subsection{Diagrammatic representation of amplitudes}

The wave function amplitudes $\langle x|\Psi\rangle$ that enter into the evaluation of the variational energy and probabilities have a relatively simple form that can be derived algebraically. In Eq.(\ref{phis}) we explicitly obtain them for configurations with up to three excitations. Although their derivation is relatively straightforward, the number of terms grows very rapidly with the number of excitations. By inspection, one can observe that the overlaps are given by the sum of all possible products of $t$'s involving the blocks $I,J,K,\cdots$ without repetition of indices, and accounting for $I<J<K<\cdots$. 
%An amplitude with $n$ excitations can be obtained by considering all the integer partitions of the number $n$, and  build
One way to represent these products is through Young tableaux. Young tableaux are partitions of the natural numbers into blocks. For example, a valid Young tableau
of size 6 is the following:

\begin{center}
\ytableausetup{textmode, boxframe=0.1em}
\begin{ytableau}
0 & 1 & 2 \\
3 & 4  \\
5
\end{ytableau}
\end{center}

We consider tableaux to be equivalent if rows are swapped.
The numbers within each row cannot be permuted, as the order of the indices must be strictly increasing.
For the set of valid unique tableaux of size $n$,
we can easily generate all valid unique tableaux
of size $n + 1$. First we start with the base case of size
one, where there is only one diagram

\begin{center}
\ytableausetup{textmode, boxframe=0.1em}
\begin{ytableau}
0 
\end{ytableau}
\end{center}

We then add the next value, 1, to the end of each
row to generate a new diagram, and finally adding
it as a new row of 1:

\begin{center}
\ytableausetup{textmode, boxframe=0.1em}
\begin{ytableau}
0 & 1
\end{ytableau} 
\begin{ytableau}
0 \\
1
\end{ytableau}
\end{center}
We can repeat this step again to get all the unique
diagrams of size 3:
\begin{center}
\ytableausetup{textmode, boxframe=0.1em}
\begin{ytableau}
0 & 1 & 2
\end{ytableau} 
\begin{ytableau}
0 & 1 \\
2
\end{ytableau}
\begin{ytableau}
0 & 2 \\
1
\end{ytableau}
\begin{ytableau}
0 \\
1 & 2
\end{ytableau}
\begin{ytableau}
0 \\
1 \\
2
\end{ytableau}
\end{center}

The size of the tableaux represents the number of excitations in a particular configuration $|x\rangle$, and each row represents one single, double, triple, etc. depending on the number of columns. 

The procedure to calculate the overlap $\langle x|\Psi\rangle$ consists of first removing all the ``zeros'' from $|00\cdots 0 i_I0 \cdots 0 j_J 0 \cdots 0 k_K\cdots 0\rangle\rightarrow |i_Ij_Jk_K\rangle$. The indices in each box of the tableaux correspond to the positions of the excitations in the compressed representation: $0\rightarrow i_I; 1\rightarrow j_J; 2\rightarrow k_K; \cdots$. As an example, for some arbitraty configuration (and counting left to right from zero)
$|010021\rangle\rightarrow |1_12_41_5\rangle$, we would use all diagrams previously
calculated for size 3 to easily generate
the following coefficients (compare to the expression in Eq.(\ref{phis})):
\[
\langle 010021|\Psi\rangle =  
t^{145}_{121}
 + t^{14}_{12}t^5_1 + t^{15}_{11}t^4_2+t^{45}_{21}t^2_1
+t^1_1 t^4_2 t^5_1
\]
This way, one can easily and efficiently generate and store all the diagrams for an arbitrary number of excitors before running the simulations, typically in a list that can be easily accessed. Moreover, one can readily calculate the derivatives with respect to the coefficients diagramatically. For instance, in the previous example:
\[
\frac{\partial \langle 010021|\Psi\rangle}{\partial t_1^5} = t_{12}^{14}+t_1^1 t_2^4.
\]
This result can be obtained by looking for the tableaux containing a row with a single box labeled as ``2'' (corresponding to $1_5$), and removing such row from the diagram.
\[
\frac{\partial \langle x|\Psi\rangle}{\partial\,
{\ytableausetup{boxsize=1em}
\begin{ytableau}
2
\end{ytableau}}
}
=
\ytableausetup{boxsize=1.5em}
\begin{ytableau}
0 & 1
\end{ytableau}
+
\begin{ytableau}
0 \\
1
\end{ytableau}
\]
Similarly, derivatives with respect to doubles are obtained by removing the row with the corresponding double excitors. 

Notice that, even though we have used examples with only three excitors, this procedure can be generalized to any order. In the case of BCCC$n$, only diagrams with at most $n$ columns per row would be kept.

The computational cost of the simulation is determined by the number of variational parameters, and the number of possible diagrams for a given number of excitations\cite{Scott2017}. In systems with translational symmetry, one can use relative distances to reduce the number of $t_2$'s to $N\times D^2$,and $t_3$'s to $N^2\times D^3$, where $N$ is the number of plaquettes, and $D$ is the local block dimension.

\subsection{Wave function optimization}

Although many methods can be used to optimize the amplitudes $t$ \cite{Sorella2000,Sorella2005,Umrigar2005,Neuscamman2011,Neuscamman2012,chen2024,Kingma2014AdamAM}, we hereby illustrate the implementation of a basic gradient descent
approach for demonstration purposes. Accordingly, the CC amplitudes are updated following the rule:
\begin{equation}
t_{\text{new}}=t_{\text{old}}-\eta \left.\frac{\partial E}{\partial t} \right|_{t_{\text{old}}},
\end{equation}
where $\eta$ is hyperparameter referred-to as the ``learning rate'', and $t$ represents an arbitrary variational parameter in the wave-function. Regardless of our choice, all optimization algorithms rely on the calculation of the partial derivatives, which can be obtained in terms of expectations values as:
\begin{eqnarray}
\frac{\partial E}{\partial t} &=& \frac{\langle \Psi|H O_t|\Psi \rangle}{\langle \Psi|\Psi \rangle} - \frac{\langle \Psi|H|\Psi \rangle}{\langle \Psi|\Psi \rangle} \frac{\langle \Psi|O_t|\Psi \rangle}{\langle \Psi|\Psi \rangle} \nonumber \\
&=& \langle H O_t \rangle-E\langle O_t \rangle.
\end{eqnarray}
In this expression, we introduce the log derivative operator\cite{Sorella2005} $O_t=\frac{1}{\Psi}\partial \Psi/\partial t$.

The numerical evaluation of these expectation values is carried out in similar fashion as the energy, Eq.(\ref{Ev}):
\[
\langle O_t \rangle = \sum_x P_x O_{t_x},
\]
with 
\[
O_{t_x}=\frac{1}{\langle x|\Psi\rangle}\frac{\partial \langle x|\Psi\rangle}{\partial t}.
\]
The derivatives of the wave-function coefficients can be readily obtained in terms of Young tableaux, previously stored in memory.

\subsection{Importance Sampling}

The Monte Carlo method relies on two fundamental conditions: detailed balance and ergodicity. These are both satisfied in principle by the use of von Neumann rejection through the definition of the update probability $w$, and by the fact that all ground state configurations are accessible by the action of the Hamiltonian $H$ through (\ref{Hx}). However, in our case, the probabilities $p_x$ are sharply peaked at $|x\rangle=|0\rangle$ because $|\Phi_0\rangle$ concentrates much of the ground state amplitude. This is more marked with larger $N_0$. Hence, most of the states $|\Phi^{i_I}\rangle,|\Phi^{i_Ij_J}\rangle,|\Phi^{i_Ij_Jk_K}\rangle,\cdots$ will have exponentially small weights that will form a long tail of fluctuations on top of the vacuum. Accessing these states through Monte Carlo sampling becomes increasingly difficult as the number of excitors grows. For practical purposes, ergodicity is effectively broken: the algorithm will not have a chance to visit most of these configurations, and the simulation becomes trapped at $|\Phi_0\rangle$ or states with very few excitations. Fortunately, there is a way to overcome this hurdle, consisting of applying importance sampling.

In importance sampling, one replaces the probability amplitude $p_x$ by a different one, $q_x$, such that
\begin{eqnarray}
E&=&\frac{1}{Z_P}\sum_x p_x E_x = \frac{Z_Q}{Z_PZ_Q}\sum_x q_x \left(\frac{p_x}{q_x} E_x\right) \nonumber \\
&=& \frac{\sum_x Q_x (w_x E_x)}{Z_P/Z_Q}, 
\label{importance1}
\end{eqnarray}
where $w_x=p_x/q_x$, $Z_P=\sum_z p_x$, $Z_Q=\sum_z q_x$, $Q_x=q_x/Z_Q$ and
\begin{equation}
\frac{Z_P}{Z_Q}=\frac{1}{Z_Q}\sum_x q_x\frac{p_x}{q_x}=\sum_x Q_x w_x.
\label{importance2}
\end{equation}
From (\ref{importance1}) and (\ref{importance2}), we find that the price we have to pay is to sample both $w_xE_x$ and $w_x$. In practice, one introduces a new hyperparameter $\alpha$, with $0<\alpha<1$, and defines a new probability $q_x=p_x^\alpha$. The hyperparameter $\alpha$ is then tuned to improve convergence, much like the learning rate\cite{misery2026importance}. A simple interpretation of this approach is to think of $\alpha$ as an analog to ``inverse temperature''. For $\alpha=0$, the configurations are sampled according to a uniform distribution, as if the system were at infinite temperature. On the other hand $\alpha=1$ corresponds to the ``true zero temperature distribution''. 

\begin{figure}
\centering
\includegraphics[width = 0.45\textwidth]{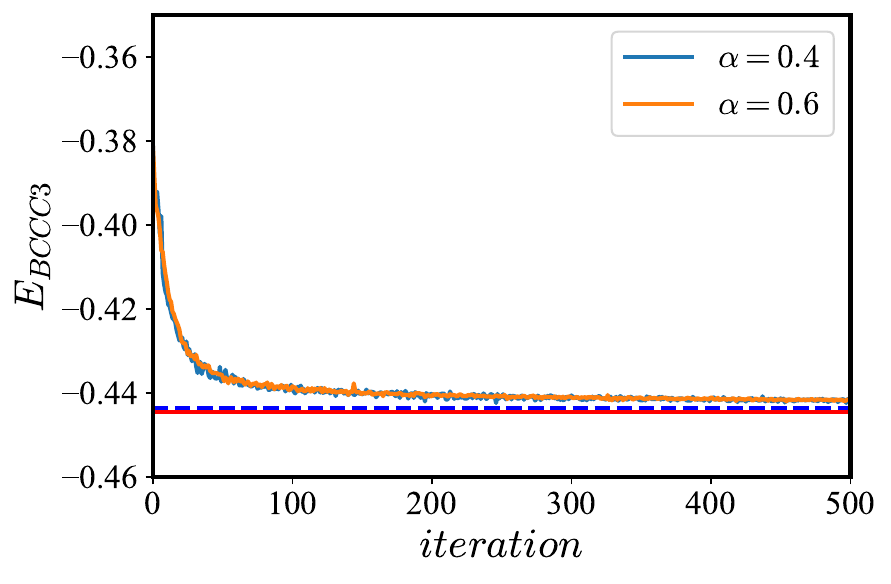}
\caption{Energy convergence for a 1D chain of length $L=24$ and block size $N_0=4$ (6 blocks) with periodic boundary conditions. We use BCCC3 and importance sampling with two different values of $\alpha$. The dashed line is the final estimate after 5000 minimization steps (we only show the first 500), while the full line is the exact result from DMRG. }
\label{fig:convergence}
\end{figure}

\begin{figure}
\centering
\includegraphics[width = 0.45\textwidth]{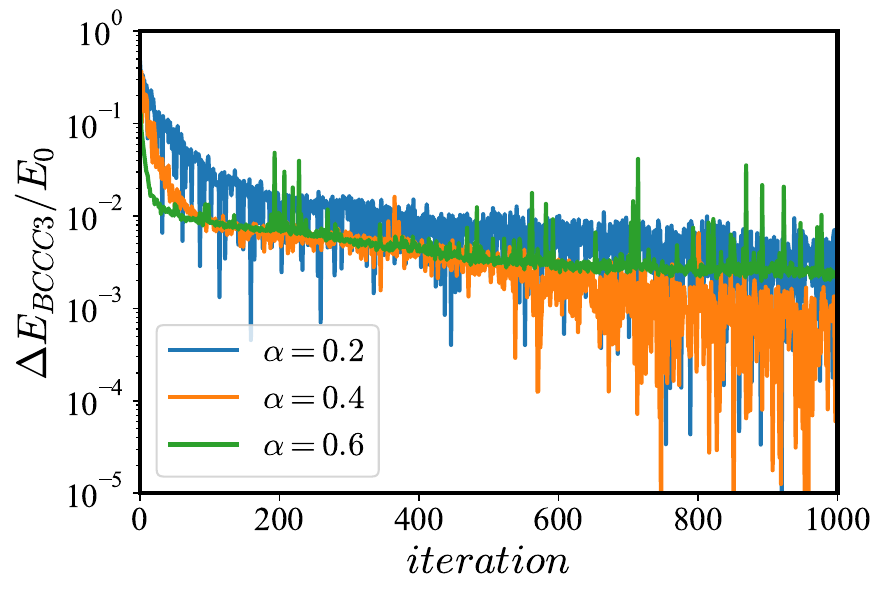}
\caption{Relative error in a log scale for a 1D chain of length $L=24$ and block size $N_0=6$ (4 blocks) with periodic boundary conditions. We use BCCC3 and importance sampling with three different values of $\alpha$.}
\label{fig:convergence_log}
\end{figure}

\section{Results}\label{sec:results}

%In order to study the Heisenberg model we first subdivide the lattice into smaller blocks of size $N_0$. The block Hamiltonian $H_0$ with open boundary conditions is subsequently diagonalized, and all eigenstates and eigenvalues calculated exactly. Our reference state is taken to be the product state of all single-plaquette ground states.  
 
We carry our simulations with 5000 optimization steps, 15000 measurements per step and leaving 10 decorrelation steps between measurements. We typically use a learning rate $\eta=0.005$. A summary of the main results can be found in Table ~\ref{tab:table1}. The convergence of a typical BCCC3 run is shown in Fig.\ref{fig:convergence}, where we simulate a chain  of length $L=24$ with periodic boundary conditions, with block size $N_0=4$ (6 blocks of size 4). The final converged energy is $E_{BCCC3}=-0.4433(5)$, compared with the exact result $E_0=-0.444535$. The accuracy can be improved by increasing the block size to $N_0=6$, paying the price of a block dimension 4 times larger. In this case, the converged energy is $E_{BCCC3}=-0.4444(3)$.

As noted previously, importance sampling becomes increasingly important as the cluster size grows. To illustrate this behavior, in Fig.\ref{fig:convergence_log} we show the relative error for the previous case with block size $N_0=6$. For small $\alpha$, the random walk exhibits larger fluctuations in both energy and derivatives, slowing the convergence. It is possible to use an adaptive procedure to determine the optimal value of $\alpha$ \cite{misery2026importance}, but here we find, by simple experimentation, the ``sweet spot'' to be around $\alpha \sim 0.4$. 

\begin{figure}
\centering
\includegraphics[width = 0.45\textwidth]{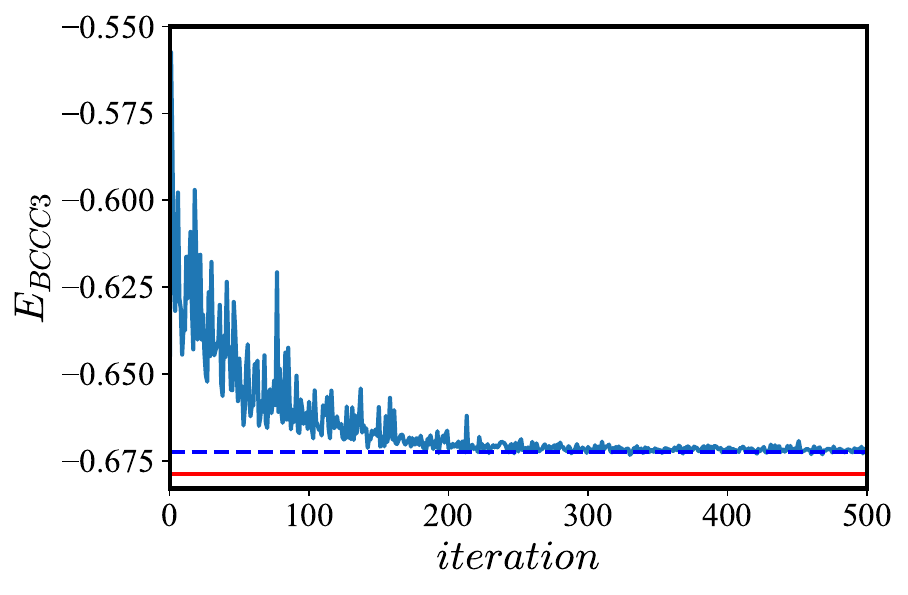}
\caption{Energy convergence for a BCCC3 simulation of a $6\times6$ square lattice using square $2\times 2$ plaquettes and periodic boundary conditions, and $\alpha=0.5$. The dashed line is the final estimate after 5000 minimization steps (we only show the first 500), while the full line is the exact result from exact diagonalization. }
\label{fig:convergence2d6}
\end{figure}

In two dimensions, we study the square lattice with blocks of size $2\times 2$.
Fig.\ref{fig:convergence2d6} shows BCCC3 the convergence for the two-dimensional $6\times6$ square lattice, using $\alpha=0.5$. Our best result is $E_{BCCC3}=-0.6725(5)$, while the exact result is $E_0=-0.67887$. 

Scaling up these simulations to larger system sizes within the full BCCC3 expansion becomes computationally prohibitive due to the extremely large number of variational parameters. For example, BCCC3 on the $8\times 8$ lattice would involve $\sim 2.5$ billion diagrams. Similarly, the total number of diagrams used in BCCC2 for the same geometry is $\sim 60$ million. %59{,}704{,}335$
%, while truncating to nine total excitors brings it down to $3{,}735$. 
Although it may be achievable using modern machine learning techniques and GPU acceleration, we hereby simply compare different truncation approaches. The BCCC2 energy for the $6 \times 6$ lattice is $E_{BCCC2}=-0.6632(3)$, indicating that the triples account for about $1.5\%$ of the correlations. For the $8\times 8$ lattice, we are forced to restrict our calculations to states with no more than nine excitors, obtaining $E_{BCCC2(9)}=-0.6338(3)$, while the exact energy is $E_{QMC}=-0.673487(4)$\cite{Sandvik1997}. The truncation to six excitors yields a much poorer variational energy $E_{BCCC3(6)}=-0.6185(3)$ indicating that the number of excitors is more important than the order of the expansion. Even if $t_2$ accounts for the largest contribution to the correlation energy, the fact that the system exhibits long-range correlations suggests that a truncation in the number of excitors is not a reliable path towards optimizing the calculation. Moreover, by comparing BCCC3(9) and BCCC2(9) we find very small differences in the energies, indicating a surprisingly small contribution from $t_3$ in this case. The variational energies deviate from the exact ones by $\sim 9\%$, at least twice as large as the result obtained using correlator product states \cite{Neuscamman2011}, providing further evidence that the main origin of the large error stems from the truncation on the number of excitors. 

To gain more insight on the limitations of the method, we carry out simulations on chains with $\Delta > 1$. In this regime, the ground state becomes gapped, and the spin-spin correlations decay exponentially with a characteristic correlation length $\xi$. As $\Delta$ increases and the correlation length becomes of the order of the cluster size, we expect improved precision in the BCCC3 approximation, as we observe in Table \ref{tab:table2}. Moreover, we also find a significant reduction in the variance, indicating that the variational ansatz can more accurately describe the many-body structure of the ground state wave function.  

Next, we consider a truncation in the number of cluster basis states following two different approaches: (i) we keep the $d$ eigenstates with the lowest energy, or (ii) we use a density matrix truncation scheme inspired by the density matrix renormalization group (DMRG) method\cite{White1992,White1993} as proposed in Ref.\onlinecite{Li2004}. We apply these ideas in 1D as follows: we first calculate the ground state of a chain with 3 blocks of size $N_0$ and periodic boundary conditions using exact diagonalization. After a partial trace over two of the blocks we obtain the reduced density matrix $\rho$ for a single block. Then, we fully diagonalize $\rho$ and we truncate the basis by keeping the $d$ eigenstates with the largest eigenvalues. The results obtained by following this scheme are shown in Table \ref{tab:table2}. A small but noticeable loss of accuracy is observed, even in the large-$\Delta$ case, but the error remains below $2\%$ in the worse case, with only 6 basis states. On the other hand, the errors for the energy truncation scheme are considerably larger (see first column in Table \ref{tab:table2}). This is not surprising, since it has already been demonstrated that the density matrix eigenvectors provide the optimal basis truncation\cite{White1992,White1993,Schollwock2005,Feiguin2013a}. 

\begin{table}[ht!]
  \begin{center}
    \caption{Variational energies for different lattice geometries and truncation level, compared to the exact results. The cluster size for the 1D case is in parenthesis. BCCCn(m) indicates that the calculation is limited to $m$ excitors. }
    \label{tab:table1}
%    \begin{tabular}{l|l|l|l|\l}
    \begin{tabular*}{\linewidth}{@{\extracolsep{\fill}} ccccllc }

      \hline \hline
      Lattice & BCCC$n$ & $\#$ diagrams & $\#$ parameters & $E_{Var}$ & $E_{Exact}$ & Error (\%)  \\     
      \hline 
      $L=24 (4)$ & BCCC3 & 234 & 11828 & -0.4433(5) & -0.444535 & 0.3 \\
      $L=24 (6)$ & BCCC3 & 22 & 256064 & -0.4444(3) & -0.444535 & 0.03\\ 
      $6\times 6$ & BCCC3 & 16310 & 32456 & -0.6725(5) & -0.67887 & 1 \\ 
      $6\times 6$ & BCCC2 & 3735 & 915 & -0.6632(3) & -0.67887 & 2.3 \\ 
    $8\times 8$ & BCCC3(9) & 16310 & 119851 & -0.6353(2) & -0.673487(4) & 6\\
    $8\times 8$ & BCCC3(6) & 234 & 119851 & -0.6212(3) & -0.673487(4) & 9\\
      $8\times 8$ & BCCC2(9) & 3735 & 1726 & -0.6338(3) & -0.673487(4) & 6\\
      \hline \hline 
    \end{tabular*}
  \end{center}
\end{table}

\begin{table}[ht!]
  \centering
  \caption{Variational energies for an anisotropic Heisenberg chain of length
  $L=24$ and block size $N_0=4$ and periodic boundary conditions, compared
  to exact results. The block basis is truncated to $d$ lowest energy eigenstates or the $d$ eigenstates of the reduced density matrix with the largest eigenvalues ($d=16$ means no truncation). The error corresponds to the $d=16$ case. }
  \label{tab:table2}

  \begin{tabular}{c|c|ccc|c|c}
    \hline\hline
    \multirow{2}{*}{$\Delta$}
%      & \multirow{2}{*}{eigenstates}
      & {eigenstates}
      & \multicolumn{3}{c|}{density matrix truncation}
      & \multirow{2}{*}{$E_{Exact}$}
      & \multirow{2}{*}{Error (\%)} \\
    \cline{2-5}
      & $d=6$ & $d=6$ & $d=8$ & $d=16$ & & \\
    \hline

    1   & -0.4342(1)  & -0.4362(2) & -0.4387(2) & -0.4433(5) & -0.444535 & 0.3  \\
    1.5 & -0.51232(4) & -0.51682(3) & -0.51953(5) & -0.52333(6) & -0.524770 & 0.3 \\
    2   & -0.60335(3) & -0.61350(4) & -0.61492(4) & -0.61664(6) & -0.617711 & 0.2 \\
    2.5 & -0.70523(3) & -0.71860(3) & -0.71944(3) & -0.72085(5) & -0.721120 & 0.04 \\

    \hline\hline
  \end{tabular}
\end{table}

\section{Summary and Conclusions}\label{sec:conclusion}

We have introduced an efficient variational Monte Carlo framework for solving the coupled-cluster equations stochastically. We demonstrated its performance on the one- and two-dimensional antiferromagnetic Heisenberg model using block correlated coupled-cluster expansions with double and triple excitations, establishing the approach as a reliable and systematically improvable route toward large-scale coupled-cluster calculations. At present, the principal limitation is the computational cost associated with evaluating the complete set of coupled-cluster diagrams, which restricts applications to double and triple excitations and to the system sizes considered here. Overcoming this bottleneck will be essential for extending the method to higher-order excitations and larger lattices. The use of ``locality''\cite{Roger1990,Zeng1998} (considering only excitations in neighboring blocks), and truncating the expansion to a finite number of excitations, may allow one to scale up these computations, by paying the price of a reduced accuracy. In fact, we have found that errors introduced by these approximations can be significant, likely due to the presence of long range correlations. We have provided evidence that this is the case by showing a significant reduction of the error for gapped systems with a finite correlation length. A truncation using the density matrix eigenvectors may provide a means to extend the CC expansion to higher order with less computational cost. %In this regard, it is not clear that these approximations may yield accurate results in systems that exhibit long-range or quasi-long-range order, such as the Heisenberg model. 

Although the goal of this work is to establish the main ideas through proof-of-concept calculations, several routes are possible to overcome the current computational bottlenecks. For example, the wave-function coefficients could be evaluated stochastically through random walks over the Young tableaux space, an approach that may be particularly well suited for GPU acceleration. Furthermore, the effective size of the variational space could potentially be reduced  by parameterizing the amplitudes $t$ using a representation with fewer variational parameters \cite{Yang2012}.

Finally, from these simulations we infer the obvious conclusion that the BCCC3 expansion is far from providing state of the art numerical accuracy for the Heisenberg model. We believe that this is due to the long-range nature of the correlations in the ground-state, that cannot be captured by this truncation. However, we speculate that it may be possible to combine these ideas with neural quantum states or Jastrow-like factors \cite{Neuscamman2011}, or perturbation theory to introduce the small corrections that can account for the missing correlations. 

The data that supports the findings of this study are available through Zenodo\cite{zenodo}.

\acknowledgements
AEF and AS are supported
by the National Science Foundation under grant No. DMR-2419109. The authors thank Fabian Ruehle and Jorge Dukelsky for useful discussions.

%\bibliography{references,dmrg,vmc,rbm}

\begin{thebibliography}{58}%
\makeatletter
\providecommand \@ifxundefined [1]{%
 \@ifx{#1\undefined}
}%
\providecommand \@ifnum [1]{%
 \ifnum #1\expandafter \@firstoftwo
 \else \expandafter \@secondoftwo
 \fi
}%
\providecommand \@ifx [1]{%
 \ifx #1\expandafter \@firstoftwo
 \else \expandafter \@secondoftwo
 \fi
}%
\providecommand \natexlab [1]{#1}%
\providecommand \enquote  [1]{``#1''}%
\providecommand \bibnamefont  [1]{#1}%
\providecommand \bibfnamefont [1]{#1}%
\providecommand \citenamefont [1]{#1}%
\providecommand \href@noop [0]{\@secondoftwo}%
\providecommand \href [0]{\begingroup \@sanitize@url \@href}%
\providecommand \@href[1]{\@@startlink{#1}\@@href}%
\providecommand \@@href[1]{\endgroup#1\@@endlink}%
\providecommand \@sanitize@url [0]{\catcode `\\12\catcode `\$12\catcode
  `\&12\catcode `\#12\catcode `\^12\catcode `\_12\catcode `\%12\relax}%
\providecommand \@@startlink[1]{}%
\providecommand \@@endlink[0]{}%
\providecommand \url  [0]{\begingroup\@sanitize@url \@url }%
\providecommand \@url [1]{\endgroup\@href {#1}{\urlprefix }}%
\providecommand \urlprefix  [0]{URL }%
\providecommand \Eprint [0]{\href }%
\providecommand \doibase [0]{https://doi.org/}%
\providecommand \selectlanguage [0]{\@gobble}%
\providecommand \bibinfo  [0]{\@secondoftwo}%
\providecommand \bibfield  [0]{\@secondoftwo}%
\providecommand \translation [1]{[#1]}%
\providecommand \BibitemOpen [0]{}%
\providecommand \bibitemStop [0]{}%
\providecommand \bibitemNoStop [0]{.\EOS\space}%
\providecommand \EOS [0]{\spacefactor3000\relax}%
\providecommand \BibitemShut  [1]{\csname bibitem#1\endcsname}%
\let\auto@bib@innerbib\@empty
%</preamble>
\bibitem [{\citenamefont {Carleo}\ and\ \citenamefont
  {Troyer}(2017)}]{Carleo2017}%
  \BibitemOpen
  \bibfield  {author} {\bibinfo {author} {\bibfnamefont {G.}~\bibnamefont
  {Carleo}}\ and\ \bibinfo {author} {\bibfnamefont {M.}~\bibnamefont
  {Troyer}},\ }\bibfield  {title} {\enquote {\bibinfo {title} {Solving the
  quantum many-body problem with artificial neural networks},}\ }\href
  {https://doi.org/10.1126/science.aag2302} {\bibfield  {journal} {\bibinfo
  {journal} {Science}\ }\textbf {\bibinfo {volume} {355}},\ \bibinfo {pages}
  {602--606} (\bibinfo {year} {2017})},\ \Eprint
  {https://arxiv.org/abs/http://science.sciencemag.org/content/355/6325/602.full.pdf}
  {http://science.sciencemag.org/content/355/6325/602.full.pdf} \BibitemShut
  {NoStop}%
\bibitem [{\citenamefont {Saito}(2017)}]{Saito2017}%
  \BibitemOpen
  \bibfield  {author} {\bibinfo {author} {\bibfnamefont {H.}~\bibnamefont
  {Saito}},\ }\bibfield  {title} {\enquote {\bibinfo {title} {Solving the
  bose–hubbard model with machine learning},}\ }\href
  {https://doi.org/10.7566/JPSJ.86.093001} {\bibfield  {journal} {\bibinfo
  {journal} {Journal of the Physical Society of Japan}\ }\textbf {\bibinfo
  {volume} {86}},\ \bibinfo {pages} {093001} (\bibinfo {year} {2017})},\
  \Eprint {https://arxiv.org/abs/https://doi.org/10.7566/JPSJ.86.093001}
  {https://doi.org/10.7566/JPSJ.86.093001} \BibitemShut {NoStop}%
\bibitem [{\citenamefont {Cai}\ and\ \citenamefont {Liu}(2018)}]{Cai2018}%
  \BibitemOpen
  \bibfield  {author} {\bibinfo {author} {\bibfnamefont {Z.}~\bibnamefont
  {Cai}}\ and\ \bibinfo {author} {\bibfnamefont {J.}~\bibnamefont {Liu}},\
  }\bibfield  {title} {\enquote {\bibinfo {title} {Approximating quantum
  many-body wave functions using artificial neural networks},}\ }\href
  {https://doi.org/10.1103/PhysRevB.97.035116} {\bibfield  {journal} {\bibinfo
  {journal} {Phys. Rev. B}\ }\textbf {\bibinfo {volume} {97}},\ \bibinfo
  {pages} {035116} (\bibinfo {year} {2018})}\BibitemShut {NoStop}%
\bibitem [{\citenamefont {Glasser}\ \emph {et~al.}(2018)\citenamefont
  {Glasser}, \citenamefont {Pancotti}, \citenamefont {August}, \citenamefont
  {Rodriguez},\ and\ \citenamefont {Cirac}}]{Glasser2018}%
  \BibitemOpen
  \bibfield  {author} {\bibinfo {author} {\bibfnamefont {I.}~\bibnamefont
  {Glasser}}, \bibinfo {author} {\bibfnamefont {N.}~\bibnamefont {Pancotti}},
  \bibinfo {author} {\bibfnamefont {M.}~\bibnamefont {August}}, \bibinfo
  {author} {\bibfnamefont {I.~D.}\ \bibnamefont {Rodriguez}},\ and\ \bibinfo
  {author} {\bibfnamefont {J.~I.}\ \bibnamefont {Cirac}},\ }\bibfield  {title}
  {\enquote {\bibinfo {title} {Neural-network quantum states, string-bond
  states, and chiral topological states},}\ }\href
  {https://doi.org/10.1103/PhysRevX.8.011006} {\bibfield  {journal} {\bibinfo
  {journal} {Phys. Rev. X}\ }\textbf {\bibinfo {volume} {8}},\ \bibinfo {pages}
  {011006} (\bibinfo {year} {2018})}\BibitemShut {NoStop}%
\bibitem [{\citenamefont {Medvidovi{\'c}}\ and\ \citenamefont
  {Moreno}(2024)}]{Medvidovic2024}%
  \BibitemOpen
  \bibfield  {author} {\bibinfo {author} {\bibfnamefont {M.}~\bibnamefont
  {Medvidovi{\'c}}}\ and\ \bibinfo {author} {\bibfnamefont {J.~R.}\
  \bibnamefont {Moreno}},\ }\bibfield  {title} {\enquote {\bibinfo {title}
  {Neural-network quantum states for many-body physics},}\ }\href@noop {}
  {\bibfield  {journal} {\bibinfo  {journal} {The European Physical Journal
  Plus}\ }\textbf {\bibinfo {volume} {139}},\ \bibinfo {pages} {631} (\bibinfo
  {year} {2024})}\BibitemShut {NoStop}%
\bibitem [{\citenamefont {Lange}\ \emph {et~al.}(2024)\citenamefont {Lange},
  \citenamefont {de~Walle}, \citenamefont {Abedinnia},\ and\ \citenamefont
  {Bohrdt}}]{Lange2024}%
  \BibitemOpen
  \bibfield  {author} {\bibinfo {author} {\bibfnamefont {H.}~\bibnamefont
  {Lange}}, \bibinfo {author} {\bibfnamefont {A.~V.}\ \bibnamefont {de~Walle}},
  \bibinfo {author} {\bibfnamefont {A.}~\bibnamefont {Abedinnia}},\ and\
  \bibinfo {author} {\bibfnamefont {A.}~\bibnamefont {Bohrdt}},\ }\bibfield
  {title} {\enquote {\bibinfo {title} {From architectures to applications: a
  review of neural quantum states},}\ }\href
  {https://doi.org/10.1088/2058-9565/ad7168} {\bibfield  {journal} {\bibinfo
  {journal} {Quantum Science and Technology}\ }\textbf {\bibinfo {volume} {9}}
  (\bibinfo {year} {2024}),\ 10.1088/2058-9565/ad7168}\BibitemShut {NoStop}%
\bibitem [{\citenamefont {Viteritti}, \citenamefont {Rende},\ and\
  \citenamefont {Becca}(2023)}]{viteritti2023vit}%
  \BibitemOpen
  \bibfield  {author} {\bibinfo {author} {\bibfnamefont {L.~L.}\ \bibnamefont
  {Viteritti}}, \bibinfo {author} {\bibfnamefont {R.}~\bibnamefont {Rende}},\
  and\ \bibinfo {author} {\bibfnamefont {F.}~\bibnamefont {Becca}},\ }\bibfield
   {title} {\enquote {\bibinfo {title} {Transformer variational wave functions
  for frustrated quantum spin systems},}\ }\href
  {https://doi.org/10.1103/PhysRevLett.130.236401} {\bibfield  {journal}
  {\bibinfo  {journal} {Phys. Rev. Lett.}\ }\textbf {\bibinfo {volume} {130}},\
  \bibinfo {pages} {236401} (\bibinfo {year} {2023})},\ \Eprint
  {https://arxiv.org/abs/2211.05504} {arXiv:2211.05504 [cond-mat.dis-nn]}
  \BibitemShut {NoStop}%
\bibitem [{\citenamefont {Sobral}, \citenamefont {Perle},\ and\ \citenamefont
  {Scheurer}(2025)}]{Sobral2025}%
  \BibitemOpen
  \bibfield  {author} {\bibinfo {author} {\bibfnamefont {J.~A.}\ \bibnamefont
  {Sobral}}, \bibinfo {author} {\bibfnamefont {M.}~\bibnamefont {Perle}},\ and\
  \bibinfo {author} {\bibfnamefont {M.~S.}\ \bibnamefont {Scheurer}},\
  }\bibfield  {title} {\enquote {\bibinfo {title} {Physics-informed
  transformers for electronic quantum states},}\ }\href
  {https://doi.org/10.1038/s41467-025-66844-z} {\bibfield  {journal} {\bibinfo
  {journal} {Nature Communications}\ }\textbf {\bibinfo {volume} {16}},\
  \bibinfo {pages} {10811} (\bibinfo {year} {2025})}\BibitemShut {NoStop}%
\bibitem [{\citenamefont {Rende}\ and\ \citenamefont
  {Viteritti}(2025)}]{rende2025queries}%
  \BibitemOpen
  \bibfield  {author} {\bibinfo {author} {\bibfnamefont {R.}~\bibnamefont
  {Rende}}\ and\ \bibinfo {author} {\bibfnamefont {L.~L.}\ \bibnamefont
  {Viteritti}},\ }\bibfield  {title} {\enquote {\bibinfo {title} {Are queries
  and keys always relevant? a case study on transformer wave functions},}\
  }\href {https://doi.org/10.1088/2632-2153/ada23a} {\bibfield  {journal}
  {\bibinfo  {journal} {Mach. Learn.: Sci. Technol.}\ }\textbf {\bibinfo
  {volume} {6}},\ \bibinfo {pages} {010501} (\bibinfo {year}
  {2025})}\BibitemShut {NoStop}%
\bibitem [{\citenamefont {Mezzacapo}\ \emph {et~al.}(2009)\citenamefont
  {Mezzacapo}, \citenamefont {Schuch}, \citenamefont {Boninsegni},\ and\
  \citenamefont {Cirac}}]{Mezzacapo2009}%
  \BibitemOpen
  \bibfield  {author} {\bibinfo {author} {\bibfnamefont {F.}~\bibnamefont
  {Mezzacapo}}, \bibinfo {author} {\bibfnamefont {N.}~\bibnamefont {Schuch}},
  \bibinfo {author} {\bibfnamefont {M.}~\bibnamefont {Boninsegni}},\ and\
  \bibinfo {author} {\bibfnamefont {J.~I.}\ \bibnamefont {Cirac}},\ }\bibfield
  {title} {\enquote {\bibinfo {title} {Ground-state properties of quantum
  many-body systems: entangled-plaquette states and variational monte carlo},}\
  }\href {https://doi.org/10.1088/1367-2630/11/8/083026} {\bibfield  {journal}
  {\bibinfo  {journal} {New Journal of Physics}\ }\textbf {\bibinfo {volume}
  {11}},\ \bibinfo {pages} {083026} (\bibinfo {year} {2009})}\BibitemShut
  {NoStop}%
\bibitem [{\citenamefont {Changlani}\ \emph {et~al.}(2009)\citenamefont
  {Changlani}, \citenamefont {Kinder}, \citenamefont {Umrigar},\ and\
  \citenamefont {Chan}}]{Changlani2009}%
  \BibitemOpen
  \bibfield  {author} {\bibinfo {author} {\bibfnamefont {H.~J.}\ \bibnamefont
  {Changlani}}, \bibinfo {author} {\bibfnamefont {J.~M.}\ \bibnamefont
  {Kinder}}, \bibinfo {author} {\bibfnamefont {C.~J.}\ \bibnamefont
  {Umrigar}},\ and\ \bibinfo {author} {\bibfnamefont {G.~K.-L.}\ \bibnamefont
  {Chan}},\ }\bibfield  {title} {\enquote {\bibinfo {title} {Approximating
  strongly correlated wave functions with correlator product states},}\ }\href
  {https://doi.org/10.1103/PhysRevB.80.245116} {\bibfield  {journal} {\bibinfo
  {journal} {Phys. Rev. B}\ }\textbf {\bibinfo {volume} {80}},\ \bibinfo
  {pages} {245116} (\bibinfo {year} {2009})}\BibitemShut {NoStop}%
\bibitem [{\citenamefont {Neuscamman}\ \emph {et~al.}(2011)\citenamefont
  {Neuscamman}, \citenamefont {Changlani}, \citenamefont {Kinder},\ and\
  \citenamefont {Chan}}]{Neuscamman2011}%
  \BibitemOpen
  \bibfield  {author} {\bibinfo {author} {\bibfnamefont {E.}~\bibnamefont
  {Neuscamman}}, \bibinfo {author} {\bibfnamefont {H.}~\bibnamefont
  {Changlani}}, \bibinfo {author} {\bibfnamefont {J.}~\bibnamefont {Kinder}},\
  and\ \bibinfo {author} {\bibfnamefont {G.~K.-L.}\ \bibnamefont {Chan}},\
  }\bibfield  {title} {\enquote {\bibinfo {title} {Nonstochastic algorithms for
  jastrow-slater and correlator product state wave functions},}\ }\href
  {https://doi.org/10.1103/PhysRevB.84.205132} {\bibfield  {journal} {\bibinfo
  {journal} {Phys. Rev. B}\ }\textbf {\bibinfo {volume} {84}},\ \bibinfo
  {pages} {205132} (\bibinfo {year} {2011})}\BibitemShut {NoStop}%
\bibitem [{\citenamefont {Isaev}, \citenamefont {Ortiz},\ and\ \citenamefont
  {Dukelsky}(2009)}]{Isaev2009}%
  \BibitemOpen
  \bibfield  {author} {\bibinfo {author} {\bibfnamefont {L.}~\bibnamefont
  {Isaev}}, \bibinfo {author} {\bibfnamefont {G.}~\bibnamefont {Ortiz}},\ and\
  \bibinfo {author} {\bibfnamefont {J.}~\bibnamefont {Dukelsky}},\ }\bibfield
  {title} {\enquote {\bibinfo {title} {Hierarchical mean-field approach to the
  ${J}_{1}\text{\ensuremath{-}}{J}_{2}$ heisenberg model on a square
  lattice},}\ }\href {https://doi.org/10.1103/PhysRevB.79.024409} {\bibfield
  {journal} {\bibinfo  {journal} {Phys. Rev. B}\ }\textbf {\bibinfo {volume}
  {79}},\ \bibinfo {pages} {024409} (\bibinfo {year} {2009})}\BibitemShut
  {NoStop}%
\bibitem [{\citenamefont {Papastathopoulos-Katsaros}\ \emph
  {et~al.}(2022{\natexlab{a}})\citenamefont {Papastathopoulos-Katsaros},
  \citenamefont {Jim{\'e}nez-Hoyos}, \citenamefont {Henderson},\ and\
  \citenamefont {Scuseria}}]{Kastaros2022}%
  \BibitemOpen
  \bibfield  {author} {\bibinfo {author} {\bibfnamefont {A.}~\bibnamefont
  {Papastathopoulos-Katsaros}}, \bibinfo {author} {\bibfnamefont {C.~A.}\
  \bibnamefont {Jim{\'e}nez-Hoyos}}, \bibinfo {author} {\bibfnamefont {T.~M.}\
  \bibnamefont {Henderson}},\ and\ \bibinfo {author} {\bibfnamefont {G.~E.}\
  \bibnamefont {Scuseria}},\ }\bibfield  {title} {\enquote {\bibinfo {title}
  {Coupled cluster and perturbation theories based on a cluster mean-field
  reference applied to strongly correlated spin systems},}\ }\href
  {https://doi.org/10.1021/acs.jctc.2c00338} {\bibfield  {journal} {\bibinfo
  {journal} {Journal of Chemical Theory and Computation}\ }\textbf {\bibinfo
  {volume} {18}},\ \bibinfo {pages} {4293--4303} (\bibinfo {year}
  {2022}{\natexlab{a}})}\BibitemShut {NoStop}%
\bibitem [{\citenamefont {Li}(2004)}]{Li2004}%
  \BibitemOpen
  \bibfield  {author} {\bibinfo {author} {\bibfnamefont {S.}~\bibnamefont
  {Li}},\ }\bibfield  {title} {\enquote {\bibinfo {title} {Block-correlated
  coupled cluster theory:the general formulation and its application to the
  antiferromagnetic heisenberg model},}\ }\href
  {https://doi.org/10.1063/1.1646355} {\bibfield  {journal} {\bibinfo
  {journal} {The Journal of Chemical Physics}\ }\textbf {\bibinfo {volume}
  {120}},\ \bibinfo {pages} {5017--5026} (\bibinfo {year} {2004})},\ \Eprint
  {https://arxiv.org/abs/https://pubs.aip.org/aip/jcp/article-pdf/120/11/5017/10855001/5017\_1\_online.pdf}
  {https://pubs.aip.org/aip/jcp/article-pdf/120/11/5017/10855001/5017\_1\_online.pdf}
  \BibitemShut {NoStop}%
\bibitem [{\citenamefont {Wang}\ \emph {et~al.}(2020)\citenamefont {Wang},
  \citenamefont {Duan}, \citenamefont {Xu}, \citenamefont {Zou},\ and\
  \citenamefont {Li}}]{Wang2020}%
  \BibitemOpen
  \bibfield  {author} {\bibinfo {author} {\bibfnamefont {Q.}~\bibnamefont
  {Wang}}, \bibinfo {author} {\bibfnamefont {M.}~\bibnamefont {Duan}}, \bibinfo
  {author} {\bibfnamefont {E.}~\bibnamefont {Xu}}, \bibinfo {author}
  {\bibfnamefont {J.}~\bibnamefont {Zou}},\ and\ \bibinfo {author}
  {\bibfnamefont {S.}~\bibnamefont {Li}},\ }\bibfield  {title} {\enquote
  {\bibinfo {title} {Describing strong correlation with block-correlated
  coupled cluster theory},}\ }\href
  {https://doi.org/10.1021/acs.jpclett.0c02117} {\bibfield  {journal} {\bibinfo
   {journal} {The Journal of Physical Chemistry Letters}\ }\textbf {\bibinfo
  {volume} {11}},\ \bibinfo {pages} {7536--7543} (\bibinfo {year}
  {2020})}\BibitemShut {NoStop}%
\bibitem [{\citenamefont {Ren}\ \emph {et~al.}(2024)\citenamefont {Ren},
  \citenamefont {Zou}, \citenamefont {Zhang}, \citenamefont {Li},\ and\
  \citenamefont {Li}}]{ren2024block}%
  \BibitemOpen
  \bibfield  {author} {\bibinfo {author} {\bibfnamefont {X.}~\bibnamefont
  {Ren}}, \bibinfo {author} {\bibfnamefont {J.}~\bibnamefont {Zou}}, \bibinfo
  {author} {\bibfnamefont {H.}~\bibnamefont {Zhang}}, \bibinfo {author}
  {\bibfnamefont {W.}~\bibnamefont {Li}},\ and\ \bibinfo {author}
  {\bibfnamefont {S.}~\bibnamefont {Li}},\ }\bibfield  {title} {\enquote
  {\bibinfo {title} {Block-correlated coupled cluster theory with up to
  four-pair correlation for accurate static correlation of strongly correlated
  systems},}\ }\href@noop {} {\bibfield  {journal} {\bibinfo  {journal} {The
  Journal of Physical Chemistry Letters}\ }\textbf {\bibinfo {volume} {15}},\
  \bibinfo {pages} {693--700} (\bibinfo {year} {2024})}\BibitemShut {NoStop}%
\bibitem [{\citenamefont {Bethe}(1931)}]{Bethe1031}%
  \BibitemOpen
  \bibfield  {author} {\bibinfo {author} {\bibfnamefont {H.}~\bibnamefont
  {Bethe}},\ }\bibfield  {title} {\enquote {\bibinfo {title} {Zur theorie der
  metalle},}\ }\href {https://doi.org/10.1007/BF01341708} {\bibfield  {journal}
  {\bibinfo  {journal} {Zeitschrift f{\"u}r Physik}\ }\textbf {\bibinfo
  {volume} {71}},\ \bibinfo {pages} {205--226} (\bibinfo {year}
  {1931})}\BibitemShut {NoStop}%
\bibitem [{\citenamefont {Caux}()}]{integrability}%
  \BibitemOpen
  \bibfield  {author} {\bibinfo {author} {\bibfnamefont {J.-S.}\ \bibnamefont
  {Caux}},\ }\href@noop {} {\enquote {\bibinfo {title} {The bethe ansatz},}\
  }\bibinfo {howpublished} {https://integrability.org/}\BibitemShut {NoStop}%
\bibitem [{\citenamefont {White}(1992)}]{White1992}%
  \BibitemOpen
  \bibfield  {author} {\bibinfo {author} {\bibfnamefont {S.~R.}\ \bibnamefont
  {White}},\ }\bibfield  {title} {\enquote {\bibinfo {title} {Density matrix
  formulation for quantum renormalization groups},}\ }\href
  {https://doi.org/10.1103/PhysRevLett.69.2863} {\bibfield  {journal} {\bibinfo
   {journal} {Phys. Rev. Lett.}\ }\textbf {\bibinfo {volume} {69}},\ \bibinfo
  {pages} {2863--2866} (\bibinfo {year} {1992})}\BibitemShut {NoStop}%
\bibitem [{\citenamefont {White}(1993)}]{White1993}%
  \BibitemOpen
  \bibfield  {author} {\bibinfo {author} {\bibfnamefont {S.~R.}\ \bibnamefont
  {White}},\ }\bibfield  {title} {\enquote {\bibinfo {title} {Density-matrix
  algorithms for quantum renormalization groups},}\ }\href
  {https://doi.org/10.1103/PhysRevB.48.10345} {\bibfield  {journal} {\bibinfo
  {journal} {Phys. Rev. B}\ }\textbf {\bibinfo {volume} {48}},\ \bibinfo
  {pages} {10345--10356} (\bibinfo {year} {1993})}\BibitemShut {NoStop}%
\bibitem [{\citenamefont {Schollw\"ock}(2005)}]{Schollwock2005}%
  \BibitemOpen
  \bibfield  {author} {\bibinfo {author} {\bibfnamefont {U.}~\bibnamefont
  {Schollw\"ock}},\ }\bibfield  {title} {\enquote {\bibinfo {title} {The
  density-matrix renormalization group},}\ }\href
  {https://doi.org/10.1103/RevModPhys.77.259} {\bibfield  {journal} {\bibinfo
  {journal} {Rev. Mod. Phys.}\ }\textbf {\bibinfo {volume} {77}},\ \bibinfo
  {pages} {259--315} (\bibinfo {year} {2005})}\BibitemShut {NoStop}%
\bibitem [{\citenamefont {Schollw\"ock}(2011)}]{Schollwock2011}%
  \BibitemOpen
  \bibfield  {author} {\bibinfo {author} {\bibfnamefont {U.}~\bibnamefont
  {Schollw\"ock}},\ }\bibfield  {title} {\enquote {\bibinfo {title} {The
  density-matrix renormalization group in the age of matrix product states},}\
  }\href {https://doi.org/https://doi.org/10.1016/j.aop.2010.09.012} {\bibfield
   {journal} {\bibinfo  {journal} {Annals of Physics}\ }\textbf {\bibinfo
  {volume} {326}},\ \bibinfo {pages} {96 -- 192} (\bibinfo {year} {2011})},\
  \bibinfo {note} {january 2011 Special Issue}\BibitemShut {NoStop}%
\bibitem [{\citenamefont {Feiguin}(2013)}]{Feiguin2013a}%
  \BibitemOpen
  \bibfield  {author} {\bibinfo {author} {\bibfnamefont {A.~E.}\ \bibnamefont
  {Feiguin}},\ }\bibfield  {title} {\enquote {\bibinfo {title} {The density
  matrix renormalization group},}\ }in\ \href@noop {} {\emph {\bibinfo
  {booktitle} {Strongly correlated systems: Numerical methods}}},\ \bibinfo
  {editor} {edited by\ \bibinfo {editor} {\bibfnamefont {A.}~\bibnamefont
  {Avella}}\ and\ \bibinfo {editor} {\bibfnamefont {F.}~\bibnamefont
  {Mancini}}}\ (\bibinfo  {publisher} {Springer, Heidelberg, Berlin},\ \bibinfo
  {year} {2013})\BibitemShut {NoStop}%
\bibitem [{\citenamefont {Sandvik}(1997)}]{Sandvik1997}%
  \BibitemOpen
  \bibfield  {author} {\bibinfo {author} {\bibfnamefont {A.~W.}\ \bibnamefont
  {Sandvik}},\ }\bibfield  {title} {\enquote {\bibinfo {title} {Finite-size
  scaling of the ground-state parameters of the two-dimensional heisenberg
  model},}\ }\href {https://doi.org/10.1103/PhysRevB.56.11678} {\bibfield
  {journal} {\bibinfo  {journal} {Phys. Rev. B}\ }\textbf {\bibinfo {volume}
  {56}},\ \bibinfo {pages} {11678--11690} (\bibinfo {year} {1997})}\BibitemShut
  {NoStop}%
\bibitem [{\citenamefont {K\"{u}mmel}(2003)}]{Kummel2003}%
  \BibitemOpen
  \bibfield  {author} {\bibinfo {author} {\bibfnamefont {H.~G.}\ \bibnamefont
  {K\"{u}mmel}},\ }\bibfield  {title} {\enquote {\bibinfo {title} {A biography
  of the coupled cluster method},}\ }\href
  {https://doi.org/10.1142/S0217979203020442} {\bibfield  {journal} {\bibinfo
  {journal} {International Journal of Modern Physics B}\ }\textbf {\bibinfo
  {volume} {17}},\ \bibinfo {pages} {5311--5325} (\bibinfo {year} {2003})},\
  \Eprint {https://arxiv.org/abs/https://doi.org/10.1142/S0217979203020442}
  {https://doi.org/10.1142/S0217979203020442} \BibitemShut {NoStop}%
\bibitem [{\citenamefont {Hagen}\ \emph {et~al.}(2010)\citenamefont {Hagen},
  \citenamefont {Papenbrock}, \citenamefont {Dean},\ and\ \citenamefont
  {Hjorth-Jensen}}]{Hagen2010}%
  \BibitemOpen
  \bibfield  {author} {\bibinfo {author} {\bibfnamefont {G.}~\bibnamefont
  {Hagen}}, \bibinfo {author} {\bibfnamefont {T.}~\bibnamefont {Papenbrock}},
  \bibinfo {author} {\bibfnamefont {D.~J.}\ \bibnamefont {Dean}},\ and\
  \bibinfo {author} {\bibfnamefont {M.}~\bibnamefont {Hjorth-Jensen}},\
  }\bibfield  {title} {\enquote {\bibinfo {title} {Ab initio coupled-cluster
  approach to nuclear structure with modern nucleon-nucleon interactions},}\
  }\href {https://doi.org/10.1103/PhysRevC.82.034330} {\bibfield  {journal}
  {\bibinfo  {journal} {Phys. Rev. C}\ }\textbf {\bibinfo {volume} {82}},\
  \bibinfo {pages} {034330} (\bibinfo {year} {2010})}\BibitemShut {NoStop}%
\bibitem [{\citenamefont {Hagen}\ and\ \citenamefont {Nam}(2012)}]{Hagen2012}%
  \BibitemOpen
  \bibfield  {author} {\bibinfo {author} {\bibfnamefont {G.}~\bibnamefont
  {Hagen}}\ and\ \bibinfo {author} {\bibfnamefont {H.~A.}\ \bibnamefont
  {Nam}},\ }\bibfield  {title} {\enquote {\bibinfo {title} {{Computational
  Aspects of Nuclear Coupled-Cluster Theory}},}\ }\href
  {https://doi.org/10.1143/PTPS.196.102} {\bibfield  {journal} {\bibinfo
  {journal} {Progress of Theoretical Physics Supplement}\ }\textbf {\bibinfo
  {volume} {196}},\ \bibinfo {pages} {102--110} (\bibinfo {year} {2012})},\
  \Eprint
  {https://arxiv.org/abs/https://academic.oup.com/ptps/article-pdf/doi/10.1143/PTPS.196.102/5339304/196-102.pdf}
  {https://academic.oup.com/ptps/article-pdf/doi/10.1143/PTPS.196.102/5339304/196-102.pdf}
  \BibitemShut {NoStop}%
\bibitem [{\citenamefont {Horoi}\ \emph {et~al.}(2007)\citenamefont {Horoi},
  \citenamefont {Gour}, \citenamefont {W\l{}och}, \citenamefont {Lodriguito},
  \citenamefont {Brown},\ and\ \citenamefont {Piecuch}}]{Horoi2007}%
  \BibitemOpen
  \bibfield  {author} {\bibinfo {author} {\bibfnamefont {M.}~\bibnamefont
  {Horoi}}, \bibinfo {author} {\bibfnamefont {J.~R.}\ \bibnamefont {Gour}},
  \bibinfo {author} {\bibfnamefont {M.}~\bibnamefont {W\l{}och}}, \bibinfo
  {author} {\bibfnamefont {M.~D.}\ \bibnamefont {Lodriguito}}, \bibinfo
  {author} {\bibfnamefont {B.~A.}\ \bibnamefont {Brown}},\ and\ \bibinfo
  {author} {\bibfnamefont {P.}~\bibnamefont {Piecuch}},\ }\bibfield  {title}
  {\enquote {\bibinfo {title} {Coupled-cluster and configuration-interaction
  calculations for heavy nuclei},}\ }\href
  {https://doi.org/10.1103/PhysRevLett.98.112501} {\bibfield  {journal}
  {\bibinfo  {journal} {Phys. Rev. Lett.}\ }\textbf {\bibinfo {volume} {98}},\
  \bibinfo {pages} {112501} (\bibinfo {year} {2007})}\BibitemShut {NoStop}%
\bibitem [{\citenamefont {Sun}, \citenamefont {Hagen},\ and\ \citenamefont
  {Papenbrock}(2023)}]{Sun2023}%
  \BibitemOpen
  \bibfield  {author} {\bibinfo {author} {\bibfnamefont {Z.~H.}\ \bibnamefont
  {Sun}}, \bibinfo {author} {\bibfnamefont {G.}~\bibnamefont {Hagen}},\ and\
  \bibinfo {author} {\bibfnamefont {T.}~\bibnamefont {Papenbrock}},\ }\bibfield
   {title} {\enquote {\bibinfo {title} {Coupled-cluster theory for strong
  entanglement in nuclei},}\ }\href
  {https://doi.org/10.1103/PhysRevC.108.014307} {\bibfield  {journal} {\bibinfo
   {journal} {Phys. Rev. C}\ }\textbf {\bibinfo {volume} {108}},\ \bibinfo
  {pages} {014307} (\bibinfo {year} {2023})}\BibitemShut {NoStop}%
\bibitem [{\citenamefont {Bartlett}\ and\ \citenamefont
  {Musia\l{}}(2007)}]{Barlett2007}%
  \BibitemOpen
  \bibfield  {author} {\bibinfo {author} {\bibfnamefont {R.~J.}\ \bibnamefont
  {Bartlett}}\ and\ \bibinfo {author} {\bibfnamefont {M.}~\bibnamefont
  {Musia\l{}}},\ }\bibfield  {title} {\enquote {\bibinfo {title}
  {Coupled-cluster theory in quantum chemistry},}\ }\href
  {https://doi.org/10.1103/RevModPhys.79.291} {\bibfield  {journal} {\bibinfo
  {journal} {Rev. Mod. Phys.}\ }\textbf {\bibinfo {volume} {79}},\ \bibinfo
  {pages} {291--352} (\bibinfo {year} {2007})}\BibitemShut {NoStop}%
\bibitem [{\citenamefont {Lyakh}\ \emph {et~al.}(2012)\citenamefont {Lyakh},
  \citenamefont {Musia{\l}}, \citenamefont {Lotrich},\ and\ \citenamefont
  {Bartlett}}]{Lyakh2012}%
  \BibitemOpen
  \bibfield  {author} {\bibinfo {author} {\bibfnamefont {D.~I.}\ \bibnamefont
  {Lyakh}}, \bibinfo {author} {\bibfnamefont {M.}~\bibnamefont {Musia{\l}}},
  \bibinfo {author} {\bibfnamefont {V.~F.}\ \bibnamefont {Lotrich}},\ and\
  \bibinfo {author} {\bibfnamefont {R.~J.}\ \bibnamefont {Bartlett}},\
  }\bibfield  {title} {\enquote {\bibinfo {title} {Multireference nature of
  chemistry: The coupled-cluster view},}\ }\href
  {https://doi.org/10.1021/cr2001417} {\bibfield  {journal} {\bibinfo
  {journal} {Chemical Reviews}\ }\textbf {\bibinfo {volume} {112}},\ \bibinfo
  {pages} {182--243} (\bibinfo {year} {2012})}\BibitemShut {NoStop}%
\bibitem [{\citenamefont {Zeng}, \citenamefont {Farnell},\ and\ \citenamefont
  {Bishop}(1998)}]{Zeng1998}%
  \BibitemOpen
  \bibfield  {author} {\bibinfo {author} {\bibfnamefont {C.}~\bibnamefont
  {Zeng}}, \bibinfo {author} {\bibfnamefont {D.~J.~J.}\ \bibnamefont
  {Farnell}},\ and\ \bibinfo {author} {\bibfnamefont {R.~F.}\ \bibnamefont
  {Bishop}},\ }\bibfield  {title} {\enquote {\bibinfo {title} {An efficient
  implementation of high-order coupled-cluster techniques applied to quantum
  magnets},}\ }\href {https://doi.org/10.1023/A:1023220222019} {\bibfield
  {journal} {\bibinfo  {journal} {Journal of Statistical Physics}\ }\textbf
  {\bibinfo {volume} {90}},\ \bibinfo {pages} {327--361} (\bibinfo {year}
  {1998})}\BibitemShut {NoStop}%
\bibitem [{\citenamefont {Bishop}\ \emph {et~al.}(2000)\citenamefont {Bishop},
  \citenamefont {Farnell}, \citenamefont {Krüger}, \citenamefont {Parkinson},
  \citenamefont {Richter},\ and\ \citenamefont {Zeng}}]{Bishop2000}%
  \BibitemOpen
  \bibfield  {author} {\bibinfo {author} {\bibfnamefont {R.~F.}\ \bibnamefont
  {Bishop}}, \bibinfo {author} {\bibfnamefont {D.~J.~J.}\ \bibnamefont
  {Farnell}}, \bibinfo {author} {\bibfnamefont {S.~E.}\ \bibnamefont
  {Krüger}}, \bibinfo {author} {\bibfnamefont {J.~B.}\ \bibnamefont
  {Parkinson}}, \bibinfo {author} {\bibfnamefont {J.}~\bibnamefont {Richter}},\
  and\ \bibinfo {author} {\bibfnamefont {C.}~\bibnamefont {Zeng}},\ }\bibfield
  {title} {\enquote {\bibinfo {title} {High-order coupled cluster method
  calculations for the ground- and excited-state properties of the spin-half
  xxz model},}\ }\href {https://doi.org/10.1088/0953-8984/12/30/317} {\bibfield
   {journal} {\bibinfo  {journal} {Journal of Physics: Condensed Matter}\
  }\textbf {\bibinfo {volume} {12}},\ \bibinfo {pages} {6887} (\bibinfo {year}
  {2000})}\BibitemShut {NoStop}%
\bibitem [{\citenamefont {Roger}\ and\ \citenamefont
  {Hetherington}(1990)}]{Roger1990}%
  \BibitemOpen
  \bibfield  {author} {\bibinfo {author} {\bibfnamefont {M.}~\bibnamefont
  {Roger}}\ and\ \bibinfo {author} {\bibfnamefont {J.~H.}\ \bibnamefont
  {Hetherington}},\ }\bibfield  {title} {\enquote {\bibinfo {title}
  {Coupled-cluster approximation for spin lattices: Application to solid
  $^{3}\mathrm{He}$},}\ }\href {https://doi.org/10.1103/PhysRevB.41.200}
  {\bibfield  {journal} {\bibinfo  {journal} {Phys. Rev. B}\ }\textbf {\bibinfo
  {volume} {41}},\ \bibinfo {pages} {200--219} (\bibinfo {year}
  {1990})}\BibitemShut {NoStop}%
\bibitem [{\citenamefont {Zinke}, \citenamefont {Schulenburg},\ and\
  \citenamefont {Richter}(2008)}]{Zinke2008}%
  \BibitemOpen
  \bibfield  {author} {\bibinfo {author} {\bibfnamefont {R.}~\bibnamefont
  {Zinke}}, \bibinfo {author} {\bibfnamefont {J.}~\bibnamefont {Schulenburg}},\
  and\ \bibinfo {author} {\bibfnamefont {J.}~\bibnamefont {Richter}},\
  }\bibfield  {title} {\enquote {\bibinfo {title} {Ground-state long-range
  order in quasi-one-dimensional heisenbergquantum antiferromagnets: high-order
  coupled-cluster calculations},}\ }\href
  {https://doi.org/10.1140/epjb/e2008-00055-7} {\bibfield  {journal} {\bibinfo
  {journal} {The European Physical Journal B}\ }\textbf {\bibinfo {volume}
  {61}},\ \bibinfo {pages} {147--152} (\bibinfo {year} {2008})}\BibitemShut
  {NoStop}%
\bibitem [{\citenamefont {Darradi}\ \emph {et~al.}(2008)\citenamefont
  {Darradi}, \citenamefont {Derzhko}, \citenamefont {Zinke}, \citenamefont
  {Schulenburg}, \citenamefont {Kr\"uger},\ and\ \citenamefont
  {Richter}}]{Darradi2008}%
  \BibitemOpen
  \bibfield  {author} {\bibinfo {author} {\bibfnamefont {R.}~\bibnamefont
  {Darradi}}, \bibinfo {author} {\bibfnamefont {O.}~\bibnamefont {Derzhko}},
  \bibinfo {author} {\bibfnamefont {R.}~\bibnamefont {Zinke}}, \bibinfo
  {author} {\bibfnamefont {J.}~\bibnamefont {Schulenburg}}, \bibinfo {author}
  {\bibfnamefont {S.~E.}\ \bibnamefont {Kr\"uger}},\ and\ \bibinfo {author}
  {\bibfnamefont {J.}~\bibnamefont {Richter}},\ }\bibfield  {title} {\enquote
  {\bibinfo {title} {Ground state phases of the spin-1/2 ${J}_{1}--{J}_{2}$
  heisenberg antiferromagnet on the square lattice: A high-order coupled
  cluster treatment},}\ }\href {https://doi.org/10.1103/PhysRevB.78.214415}
  {\bibfield  {journal} {\bibinfo  {journal} {Phys. Rev. B}\ }\textbf {\bibinfo
  {volume} {78}},\ \bibinfo {pages} {214415} (\bibinfo {year}
  {2008})}\BibitemShut {NoStop}%
\bibitem [{\citenamefont {Richter}\ \emph {et~al.}(2010)\citenamefont
  {Richter}, \citenamefont {Darradi}, \citenamefont {Schulenburg},
  \citenamefont {Farnell},\ and\ \citenamefont {Rosner}}]{Richter2010}%
  \BibitemOpen
  \bibfield  {author} {\bibinfo {author} {\bibfnamefont {J.}~\bibnamefont
  {Richter}}, \bibinfo {author} {\bibfnamefont {R.}~\bibnamefont {Darradi}},
  \bibinfo {author} {\bibfnamefont {J.}~\bibnamefont {Schulenburg}}, \bibinfo
  {author} {\bibfnamefont {D.~J.~J.}\ \bibnamefont {Farnell}},\ and\ \bibinfo
  {author} {\bibfnamefont {H.}~\bibnamefont {Rosner}},\ }\bibfield  {title}
  {\enquote {\bibinfo {title} {Frustrated spin-$\frac{1}{2}$
  ${J}_{1}\text{\ensuremath{-}}{J}_{2}$ heisenberg ferromagnet on the square
  lattice studied via exact diagonalization and coupled-cluster method},}\
  }\href {https://doi.org/10.1103/PhysRevB.81.174429} {\bibfield  {journal}
  {\bibinfo  {journal} {Phys. Rev. B}\ }\textbf {\bibinfo {volume} {81}},\
  \bibinfo {pages} {174429} (\bibinfo {year} {2010})}\BibitemShut {NoStop}%
\bibitem [{\citenamefont {Richter}, \citenamefont {Zinke},\ and\ \citenamefont
  {Farnell}(2015)}]{Richter2015}%
  \BibitemOpen
  \bibfield  {author} {\bibinfo {author} {\bibfnamefont {J.}~\bibnamefont
  {Richter}}, \bibinfo {author} {\bibfnamefont {R.}~\bibnamefont {Zinke}},\
  and\ \bibinfo {author} {\bibfnamefont {D.~J.~J.}\ \bibnamefont {Farnell}},\
  }\bibfield  {title} {\enquote {\bibinfo {title} {The spin-1/2 square-lattice
  j1-j2 model: the spin-gap issue},}\ }\href
  {https://doi.org/10.1140/epjb/e2014-50589-x} {\bibfield  {journal} {\bibinfo
  {journal} {The European Physical Journal B}\ }\textbf {\bibinfo {volume}
  {88}},\ \bibinfo {pages} {2} (\bibinfo {year} {2015})}\BibitemShut {NoStop}%
\bibitem [{\citenamefont {Bishop}\ \emph {et~al.}(2019)\citenamefont {Bishop},
  \citenamefont {Li}, \citenamefont {G\"otze},\ and\ \citenamefont
  {Richter}}]{Bishop2019}%
  \BibitemOpen
  \bibfield  {author} {\bibinfo {author} {\bibfnamefont {R.~F.}\ \bibnamefont
  {Bishop}}, \bibinfo {author} {\bibfnamefont {P.~H.~Y.}\ \bibnamefont {Li}},
  \bibinfo {author} {\bibfnamefont {O.}~\bibnamefont {G\"otze}},\ and\ \bibinfo
  {author} {\bibfnamefont {J.}~\bibnamefont {Richter}},\ }\bibfield  {title}
  {\enquote {\bibinfo {title} {Frustrated spin-$\frac{1}{2}$ heisenberg magnet
  on a square-lattice bilayer: High-order study of the quantum critical
  behavior of the
  ${J}_{1}\text{\ensuremath{-}}{J}_{2}\text{\ensuremath{-}}{J}_{1}^{\ensuremath{\perp}}$
  model},}\ }\href {https://doi.org/10.1103/PhysRevB.100.024401} {\bibfield
  {journal} {\bibinfo  {journal} {Phys. Rev. B}\ }\textbf {\bibinfo {volume}
  {100}},\ \bibinfo {pages} {024401} (\bibinfo {year} {2019})}\BibitemShut
  {NoStop}%
\bibitem [{\citenamefont {Papastathopoulos-Katsaros}\ \emph
  {et~al.}(2022{\natexlab{b}})\citenamefont {Papastathopoulos-Katsaros},
  \citenamefont {Jim{\'e}nez-Hoyos}, \citenamefont {Henderson},\ and\
  \citenamefont {Scuseria}}]{Papas2022}%
  \BibitemOpen
  \bibfield  {author} {\bibinfo {author} {\bibfnamefont {A.}~\bibnamefont
  {Papastathopoulos-Katsaros}}, \bibinfo {author} {\bibfnamefont {C.~A.}\
  \bibnamefont {Jim{\'e}nez-Hoyos}}, \bibinfo {author} {\bibfnamefont {T.~M.}\
  \bibnamefont {Henderson}},\ and\ \bibinfo {author} {\bibfnamefont {G.~E.}\
  \bibnamefont {Scuseria}},\ }\bibfield  {title} {\enquote {\bibinfo {title}
  {Coupled cluster and perturbation theories based on a cluster mean-field
  reference applied to strongly correlated spin systems},}\ }\href
  {https://doi.org/10.1021/acs.jctc.2c00338} {\bibfield  {journal} {\bibinfo
  {journal} {Journal of Chemical Theory and Computation}\ }\textbf {\bibinfo
  {volume} {18}},\ \bibinfo {pages} {4293--4303} (\bibinfo {year}
  {2022}{\natexlab{b}})}\BibitemShut {NoStop}%
\bibitem [{\citenamefont {Thom}(2010)}]{Thom2010}%
  \BibitemOpen
  \bibfield  {author} {\bibinfo {author} {\bibfnamefont {A.~J.~W.}\
  \bibnamefont {Thom}},\ }\bibfield  {title} {\enquote {\bibinfo {title}
  {Stochastic coupled cluster theory},}\ }\href
  {https://doi.org/10.1103/PhysRevLett.105.263004} {\bibfield  {journal}
  {\bibinfo  {journal} {Phys. Rev. Lett.}\ }\textbf {\bibinfo {volume} {105}},\
  \bibinfo {pages} {263004} (\bibinfo {year} {2010})}\BibitemShut {NoStop}%
\bibitem [{\citenamefont {Scott}\ and\ \citenamefont {Thom}(2017)}]{Scott2017}%
  \BibitemOpen
  \bibfield  {author} {\bibinfo {author} {\bibfnamefont {C.~J.~C.}\
  \bibnamefont {Scott}}\ and\ \bibinfo {author} {\bibfnamefont {A.~J.~W.}\
  \bibnamefont {Thom}},\ }\bibfield  {title} {\enquote {\bibinfo {title}
  {{Stochastic coupled cluster theory: Efficient sampling of the coupled
  cluster expansion}},}\ }\href {https://doi.org/10.1063/1.4991795} {\bibfield
  {journal} {\bibinfo  {journal} {The Journal of Chemical Physics}\ }\textbf
  {\bibinfo {volume} {147}},\ \bibinfo {pages} {124105} (\bibinfo {year}
  {2017})},\ \Eprint
  {https://arxiv.org/abs/https://pubs.aip.org/aip/jcp/article-pdf/doi/10.1063/1.4991795/15533537/124105\_1\_online.pdf}
  {https://pubs.aip.org/aip/jcp/article-pdf/doi/10.1063/1.4991795/15533537/124105\_1\_online.pdf}
  \BibitemShut {NoStop}%
\bibitem [{\citenamefont {Spencer}\ and\ \citenamefont
  {Thom}(2016)}]{Spencer2016}%
  \BibitemOpen
  \bibfield  {author} {\bibinfo {author} {\bibfnamefont {J.~S.}\ \bibnamefont
  {Spencer}}\ and\ \bibinfo {author} {\bibfnamefont {A.~J.~W.}\ \bibnamefont
  {Thom}},\ }\bibfield  {title} {\enquote {\bibinfo {title} {{Developments in
  stochastic coupled cluster theory: The initiator approximation and
  application to the uniform electron gas}},}\ }\href
  {https://doi.org/10.1063/1.4942173} {\bibfield  {journal} {\bibinfo
  {journal} {The Journal of Chemical Physics}\ }\textbf {\bibinfo {volume}
  {144}},\ \bibinfo {pages} {084108} (\bibinfo {year} {2016})},\ \Eprint
  {https://arxiv.org/abs/https://pubs.aip.org/aip/jcp/article-pdf/doi/10.1063/1.4942173/15509820/084108\_1\_online.pdf}
  {https://pubs.aip.org/aip/jcp/article-pdf/doi/10.1063/1.4942173/15509820/084108\_1\_online.pdf}
  \BibitemShut {NoStop}%
\bibitem [{\citenamefont {Crawford}\ and\ \citenamefont
  {Schaefer~III}(2000)}]{Crawford2000}%
  \BibitemOpen
  \bibfield  {author} {\bibinfo {author} {\bibfnamefont {T.~D.}\ \bibnamefont
  {Crawford}}\ and\ \bibinfo {author} {\bibfnamefont {H.~F.}\ \bibnamefont
  {Schaefer~III}},\ }\enquote {\bibinfo {title} {An introduction to coupled
  cluster theory for computational chemists},}\ in\ \href
  {https://doi.org/https://doi.org/10.1002/9780470125915.ch2} {\emph {\bibinfo
  {booktitle} {Reviews in Computational Chemistry}}}\ (\bibinfo  {publisher}
  {John Wiley $\&$ Sons, Ltd},\ \bibinfo {year} {2000})\ pp.\ \bibinfo {pages}
  {33--136},\ \Eprint
  {https://arxiv.org/abs/https://onlinelibrary.wiley.com/doi/pdf/10.1002/9780470125915.ch2}
  {https://onlinelibrary.wiley.com/doi/pdf/10.1002/9780470125915.ch2}
  \BibitemShut {NoStop}%
\bibitem [{\citenamefont {Raghavachari}\ \emph {et~al.}(1989)\citenamefont
  {Raghavachari}, \citenamefont {Trucks}, \citenamefont {Pople},\ and\
  \citenamefont {Head-Gordon}}]{RAGHAVACHARI1989}%
  \BibitemOpen
  \bibfield  {author} {\bibinfo {author} {\bibfnamefont {K.}~\bibnamefont
  {Raghavachari}}, \bibinfo {author} {\bibfnamefont {G.~W.}\ \bibnamefont
  {Trucks}}, \bibinfo {author} {\bibfnamefont {J.~A.}\ \bibnamefont {Pople}},\
  and\ \bibinfo {author} {\bibfnamefont {M.}~\bibnamefont {Head-Gordon}},\
  }\bibfield  {title} {\enquote {\bibinfo {title} {A fifth-order perturbation
  comparison of electron correlation theories},}\ }\href
  {https://doi.org/https://doi.org/10.1016/S0009-2614(89)87395-6} {\bibfield
  {journal} {\bibinfo  {journal} {Chemical Physics Letters}\ }\textbf {\bibinfo
  {volume} {157}},\ \bibinfo {pages} {479--483} (\bibinfo {year}
  {1989})}\BibitemShut {NoStop}%
\bibitem [{\citenamefont {Noga}\ and\ \citenamefont
  {Bartlett}(1987)}]{Noga1987}%
  \BibitemOpen
  \bibfield  {author} {\bibinfo {author} {\bibfnamefont {J.}~\bibnamefont
  {Noga}}\ and\ \bibinfo {author} {\bibfnamefont {R.~J.}\ \bibnamefont
  {Bartlett}},\ }\bibfield  {title} {\enquote {\bibinfo {title} {The full ccsdt
  model for molecular electronic structure},}\ }\href
  {https://doi.org/10.1063/1.452353} {\bibfield  {journal} {\bibinfo  {journal}
  {The Journal of Chemical Physics}\ }\textbf {\bibinfo {volume} {86}},\
  \bibinfo {pages} {7041--7050} (\bibinfo {year} {1987})}\BibitemShut {NoStop}%
\bibitem [{\citenamefont {Arponen}(1983)}]{ARPONEN1983311}%
  \BibitemOpen
  \bibfield  {author} {\bibinfo {author} {\bibfnamefont {J.}~\bibnamefont
  {Arponen}},\ }\bibfield  {title} {\enquote {\bibinfo {title} {Variational
  principles and linked-cluster exp s expansions for static and dynamic
  many-body problems},}\ }\href
  {https://doi.org/https://doi.org/10.1016/0003-4916(83)90284-1} {\bibfield
  {journal} {\bibinfo  {journal} {Annals of Physics}\ }\textbf {\bibinfo
  {volume} {151}},\ \bibinfo {pages} {311--382} (\bibinfo {year}
  {1983})}\BibitemShut {NoStop}%
\bibitem [{\citenamefont {Bartlett}\ and\ \citenamefont
  {Noga}(1988)}]{BARTLETT198829}%
  \BibitemOpen
  \bibfield  {author} {\bibinfo {author} {\bibfnamefont {R.~J.}\ \bibnamefont
  {Bartlett}}\ and\ \bibinfo {author} {\bibfnamefont {J.}~\bibnamefont
  {Noga}},\ }\bibfield  {title} {\enquote {\bibinfo {title} {The expectation
  value coupled-cluster method and analytical energy derivatives},}\ }\href
  {https://doi.org/https://doi.org/10.1016/0009-2614(88)80392-0} {\bibfield
  {journal} {\bibinfo  {journal} {Chemical Physics Letters}\ }\textbf {\bibinfo
  {volume} {150}},\ \bibinfo {pages} {29--36} (\bibinfo {year}
  {1988})}\BibitemShut {NoStop}%
\bibitem [{\citenamefont {Sorella}\ and\ \citenamefont
  {Capriotti}(2000)}]{Sorella2000}%
  \BibitemOpen
  \bibfield  {author} {\bibinfo {author} {\bibfnamefont {S.}~\bibnamefont
  {Sorella}}\ and\ \bibinfo {author} {\bibfnamefont {L.}~\bibnamefont
  {Capriotti}},\ }\bibfield  {title} {\enquote {\bibinfo {title} {{Green
  function Monte Carlo with stochastic reconfiguration: An effective remedy for
  the sign problem}},}\ }\href@noop {} {\bibfield  {journal} {\bibinfo
  {journal} {Phys. Rev. B}\ }\textbf {\bibinfo {volume} {61}},\ \bibinfo
  {pages} {2599--2612} (\bibinfo {year} {2000})}\BibitemShut {NoStop}%
\bibitem [{\citenamefont {Sorella}(2005)}]{Sorella2005}%
  \BibitemOpen
  \bibfield  {author} {\bibinfo {author} {\bibfnamefont {S.}~\bibnamefont
  {Sorella}},\ }\bibfield  {title} {\enquote {\bibinfo {title} {Wave function
  optimization in the variational monte carlo method},}\ }\href
  {https://doi.org/10.1103/PhysRevB.71.241103} {\bibfield  {journal} {\bibinfo
  {journal} {Phys. Rev. B}\ }\textbf {\bibinfo {volume} {71}},\ \bibinfo
  {pages} {241103} (\bibinfo {year} {2005})}\BibitemShut {NoStop}%
\bibitem [{\citenamefont {Umrigar}\ and\ \citenamefont
  {Filippi}(2005)}]{Umrigar2005}%
  \BibitemOpen
  \bibfield  {author} {\bibinfo {author} {\bibfnamefont {C.~J.}\ \bibnamefont
  {Umrigar}}\ and\ \bibinfo {author} {\bibfnamefont {C.}~\bibnamefont
  {Filippi}},\ }\bibfield  {title} {\enquote {\bibinfo {title} {Energy and
  variance optimization of many-body wave functions},}\ }\href
  {https://doi.org/10.1103/PhysRevLett.94.150201} {\bibfield  {journal}
  {\bibinfo  {journal} {Phys. Rev. Lett.}\ }\textbf {\bibinfo {volume} {94}},\
  \bibinfo {pages} {150201} (\bibinfo {year} {2005})}\BibitemShut {NoStop}%
\bibitem [{\citenamefont {Neuscamman}, \citenamefont {Umrigar},\ and\
  \citenamefont {Chan}(2012)}]{Neuscamman2012}%
  \BibitemOpen
  \bibfield  {author} {\bibinfo {author} {\bibfnamefont {E.}~\bibnamefont
  {Neuscamman}}, \bibinfo {author} {\bibfnamefont {C.~J.}\ \bibnamefont
  {Umrigar}},\ and\ \bibinfo {author} {\bibfnamefont {G.~K.-L.}\ \bibnamefont
  {Chan}},\ }\bibfield  {title} {\enquote {\bibinfo {title} {Optimizing large
  parameter sets in variational quantum monte carlo},}\ }\href
  {https://doi.org/10.1103/PhysRevB.85.045103} {\bibfield  {journal} {\bibinfo
  {journal} {Phys. Rev. B}\ }\textbf {\bibinfo {volume} {85}},\ \bibinfo
  {pages} {045103} (\bibinfo {year} {2012})}\BibitemShut {NoStop}%
\bibitem [{\citenamefont {Chen}\ and\ \citenamefont {Heyl}(2024)}]{chen2024}%
  \BibitemOpen
  \bibfield  {author} {\bibinfo {author} {\bibfnamefont {A.}~\bibnamefont
  {Chen}}\ and\ \bibinfo {author} {\bibfnamefont {M.}~\bibnamefont {Heyl}},\
  }\bibfield  {title} {\enquote {\bibinfo {title} {Empowering deep neural
  quantum states through efficient optimization},}\ }\href
  {https://doi.org/10.1038/s41567-024-02566-1} {\bibfield  {journal} {\bibinfo
  {journal} {Nature Physics}\ }\textbf {\bibinfo {volume} {20}},\ \bibinfo
  {pages} {1476--1481} (\bibinfo {year} {2024})}\BibitemShut {NoStop}%
\bibitem [{\citenamefont {Kingma}\ and\ \citenamefont
  {Ba}(2014)}]{Kingma2014AdamAM}%
  \BibitemOpen
  \bibfield  {author} {\bibinfo {author} {\bibfnamefont {D.~P.}\ \bibnamefont
  {Kingma}}\ and\ \bibinfo {author} {\bibfnamefont {J.}~\bibnamefont {Ba}},\
  }\bibfield  {title} {\enquote {\bibinfo {title} {Adam: A method for
  stochastic optimization},}\ }\href
  {https://api.semanticscholar.org/CorpusID:6628106} {\bibfield  {journal}
  {\bibinfo  {journal} {CoRR}\ }\textbf {\bibinfo {volume} {abs/1412.6980}}
  (\bibinfo {year} {2014})}\BibitemShut {NoStop}%
\bibitem [{\citenamefont {Misery}\ \emph {et~al.}(2026)\citenamefont {Misery},
  \citenamefont {Gravina}, \citenamefont {Santini},\ and\ \citenamefont
  {Vicentini}}]{misery2026importance}%
  \BibitemOpen
  \bibfield  {author} {\bibinfo {author} {\bibfnamefont {A.}~\bibnamefont
  {Misery}}, \bibinfo {author} {\bibfnamefont {L.}~\bibnamefont {Gravina}},
  \bibinfo {author} {\bibfnamefont {A.}~\bibnamefont {Santini}},\ and\ \bibinfo
  {author} {\bibfnamefont {F.}~\bibnamefont {Vicentini}},\ }\bibfield  {title}
  {\enquote {\bibinfo {title} {Looking elsewhere: Improving variational monte
  carlo gradients by importance sampling},}\ }\href
  {https://doi.org/10.1088/2632-2153/ae387f} {\bibfield  {journal} {\bibinfo
  {journal} {Mach. Learn.: Sci. Technol.}\ }\textbf {\bibinfo {volume} {7}},\
  \bibinfo {pages} {015035} (\bibinfo {year} {2026})},\ \Eprint
  {https://arxiv.org/abs/2507.05352} {arXiv:2507.05352 [quant-ph]} \BibitemShut
  {NoStop}%
\bibitem [{\citenamefont {Yang}\ \emph {et~al.}(2012)\citenamefont {Yang},
  \citenamefont {Chan}, \citenamefont {Manby}, \citenamefont {Sch{\"u}tz},\
  and\ \citenamefont {Werner}}]{Yang2012}%
  \BibitemOpen
  \bibfield  {author} {\bibinfo {author} {\bibfnamefont {J.}~\bibnamefont
  {Yang}}, \bibinfo {author} {\bibfnamefont {G.~K.-L.}\ \bibnamefont {Chan}},
  \bibinfo {author} {\bibfnamefont {F.~R.}\ \bibnamefont {Manby}}, \bibinfo
  {author} {\bibfnamefont {M.}~\bibnamefont {Sch{\"u}tz}},\ and\ \bibinfo
  {author} {\bibfnamefont {H.-J.}\ \bibnamefont {Werner}},\ }\bibfield  {title}
  {\enquote {\bibinfo {title} {The orbital-specific-virtual local coupled
  cluster singles and doubles method},}\ }\href
  {https://doi.org/10.1063/1.3696963} {\bibfield  {journal} {\bibinfo
  {journal} {The Journal of Chemical Physics}\ }\textbf {\bibinfo {volume}
  {136}},\ \bibinfo {pages} {144105} (\bibinfo {year} {2012})}\BibitemShut
  {NoStop}%
\bibitem [{\citenamefont {Hunter~Brody}\ and\ \citenamefont
  {Feiguin}(2026)}]{zenodo}%
  \BibitemOpen
  \bibfield  {author} {\bibinfo {author} {\bibfnamefont {A.~S.}\ \bibnamefont
  {Hunter~Brody}}\ and\ \bibinfo {author} {\bibfnamefont {A.}~\bibnamefont
  {Feiguin}},\ }\bibfield  {title} {\enquote {\bibinfo {title} {Variational
  block correlated coupled cluster monte carlo},}\ }\href
  {https://doi.org/10.5281/zenodo.21287272} {10.5281/zenodo.21287272} (\bibinfo
  {year} {2026})\BibitemShut {NoStop}%
\end{thebibliography}

%aipnum4-2.bst 2019-01-14 (MD) hand-edited version of apsrev4-1.bst

\end{document}